\documentclass[
  ,draft            
  ,numberedheadings 
  ]
  {aipproc}
\usepackage{epsfig,amssymb,amsbsy,xspace} 

\layoutstyle{6x9}

     \newcommand{\ds}{\displaystyle}
     
     \newcommand{\pathnow}{}
\newcommand{\myfig}[7]{%
\begin{figure}[#7]
\vskip #5cm	\centerline{\hspace*{#1cm}
\epsfig{width=#2cm,angle=#4,figure=\pathnow #3.ps}
	}\vskip #6cm
                    }
\newcommand{\capt}[3]%
{\caption[{#2}]{\label{Fig:#1}#3}
}

\def\qgp{quark--gluon plasma\xspace}
\def\QGP{\qgp}

\newcommand{\req}[1]{Eq.\,(\ref{eq:#1})%
}

\newcommand{\beql}[1]{
	\begin{equation} \label{eq:#1}}

\newcommand{\beqarl}[1]{
	\begin{eqnarray} \label{eq:#1} }
\newcommand{\eeql}[1]{\label{eq:#1} \end{equation} 
} 
\newcommand{\eeqarl}[1]{\label{eq:#1} \end{eqnarray} 
}

\def\beq{\begin{equation}}
\def\eeq{\end{equation}}

\def\beqar{\begin{eqnarray}}
\def\eeqar{\end{eqnarray}}
\def\bcite{\cite}
\def\agev{{$A$~GeV}\xspace}

\newcommand{\captt}[3]%
{\caption[{#2}]{\label{Tab:#1}#3%
}%
}

\newcommand{\lssec}[1]{\label{ssec:#1} 
 }

\def\ie{{i.e.\xspace}}
\def\eg{{e.g.\xspace}}

\newcommand{\myfigd}[9]{%
\begin{figure}[#9]
\vskip #5cm	\centerline{\hspace*{#1cm}
\psfig{width=#2cm,angle=#4,figure=\pathnow #3.ps}
\hspace*{#7cm}
\psfig{width=#2cm,angle=#4,figure=\pathnow #8.ps}
	}\vskip #6cm
                    }

\begin{document}

\title{Non-equilibrium Hadrochemistry in\\ QGP Hadronization}

\author{Johann Rafelski}{
  address={Department of Physics, University of Arizona, 
        Tucson, AZ, 85721}
}

\author{Jean Letessier}{
  address={Laboratoire de Physique Th\'eorique et Hautes Energies\\
Universit\'e Paris 7, 2 place Jussieu, F--75251 Cedex 05}
}

\begin{abstract}
This survey offers an introductory tutorial for students of any age 
of the currently thriving field of hadrochemistry. We discuss the 
different chemical potentials, 
how the hadronic  phase space is described and how one evaluates the
abundance of hadrons at time of hadronization. We show that a rather 
accurate description of experimental data arises and we present results
of fits to hadron yields at SPS and RHIC. We show that introduction
of chemical non-equilibrium originating in a sudden  hadronization of
a QGP is favored strongly at SPS and is presently also emerging at  RHIC. 
The low chemical freeze-out temperatures are consistent with the picture of
single freeze-out scenario (chemical and thermal freeze-out coincide). 
\end{abstract}

\maketitle


\section{Thermal equilibrium}\label{Kin}
Hadronic interactions are strong, collisions are frequent,  
and along with Hagedorn \bcite{Hag65,Hag68,Hag73},
we expect formation of statistical equilibrium conditions.  
The formation of a space--time-localized fireball of
dense matter is the  key physical process occurring in 
high energy nuclear collisions. This said,  the 
question is how  
such a fireball can possibly  arise from a rather short sequence of 
individual reactions that occur when two, rather small, 
gas clouds of partons, clustered in 
nucleons, bound in  the nucleus, collide? Indeed, at first sight, one would 
be led to believe that the small clouds comprising point-like  objects would 
mutually disperse in the collision, and no localized,  
dense state of hadronic matter should  be formed. It was 
suggested in some early work, that the two colliding `eggs' should emerge 
from the high-energy interaction slightly `warmed', but still 
largely `unbroken'. In past few years some of
 our colleagues have made these eggs not only hot but also
scrambled: they believe that somehow this hadronic fireball also 
develops a high degree of not only (kinetic) thermal, but also chemical
equilibration. What does this mean precisely, and can this be true,
is what we are going to study in depth here.

Two remarkable 
properties of hadronic interactions are responsible for
 deeply inelastic behavior, which could lead to 
localization of energy in space--time:
\begin{itemize}
\item the multiparticle production in hadron--hadron collisions; and
\item the effective size of all hadrons expressed in term of their reaction
cross sections.
\end{itemize}
What appears to be a thin system of point-like constituents is effectively 
already  a volume-filling nucleon liquid, which will undergo, in a collision, a 
rapid self-multiplication with particle density rising and individual 
scattering times becoming progressively much shorter than the overall 
collision time. In historical context, the abundant particle production
 seen in high energy cosmic ray interactions, has lead Fermi to propose 
the statistical model of particle production \cite{Fer50}. Our
present developments are a natural elaboration of this seminal
work, toward a diversified field of many flavored hadron yields, 
allowing  a precision level that
a one parameter Fermi-model could of course not reach. Our objective
is to account for rapid chemical non-equilibrium dynamics related to 
formation and sudden disintegration of a new phase of matter. 

As the energy available in the collision is increased,
 the hadron particle/energy density will reach
values at which the dissolution of the hadronic constituents into a common
deconfined domain will become possible, and indeed must occur according to 
our knowledge about strong interactions. We do not really know whether
deconfinement of hadrons is operating already at 
AGS energies \cite{Let94b}, but there is today 
no experimental evidence that this low
energy range suffices. In contradistinction, a significant 
number of results obtained at the top SPS energy range can be most naturally 
interpreted in terms of  the
formation of a deconfined space--time domain, and this
becomes a much more convincing situation once RHIC results
are considered. We note that, per participant, 
there are as many as 7--10 further hadrons produced at SPS energies.
This implies that there are thousands of quarks and gluons in the space--time
domain of interest, and hence  consideration of a `local' (in space--time)
equilibrium makes good sense. However, we need to establish 
which equilibrium is reached, and if not fully, how close are we, and
what the deviations from equilibrium tell us about the physics questions
we are studying.

In order to establish locally in space and time thermal 
equilibrium, a rapid equipartition of 
energy among the different particles present has to occur.
Thermal equilibrium can be achieved in principle solely by 
elastic scattering. Introduction  of a (local) temperature $T$ 
presupposes that thermal equilibrium has nearly been established.
What is the mechanism for the establishment of 
 kinetic momentum distribution  equilibrium? The thermalization of the momentum 
distributions is driven by {\em all} scattering processes, elastic as 
well as inelastic, because all of them are associated with  
exchange of momentum and energy between particles. 
The  {\em scattering time}, for particles of 
species $i$, is given in terms of the
 reaction cross section $\sigma_{ij}$ with particle  species $j$,
 \beql{tausc}
   \tau_{i,{\mathrm{scatt}}} 
     = {1\over \sum_j \langle \sigma_{ij} v_{ij} \rangle \rho_j}.
 \eeq
The sum in the denominator is 
over all available particle species with densities $\rho_j$, 
$v_{ij}$ are the  relative velocities,
and the average is to be taken over the 
momentum distributions of the  particle considered.  

It is not hard to `guesstimate' the time  scale
governing the kinetic equilibration in
the QGP.  The typical particle-collision time (the inverse of the collision
frequency) is obtained from \req{tausc} above. 
Given the particle densities  and soft reaction cross sections,
with  the relative velocity of these
essentially massless components being the velocity of light $c$,  we find
for the \qgp scattering time,
\beql{QGPscatti}
\tau_i^{\mathrm{QGP}}=0.2\mbox{--}2\,\mbox{fm},\,\quad \mbox{with}\ 
\rho_i=2\mbox{--}10\, \mbox{fm}^{-3}, \ 
\sigma_i=2\mbox{--}5\,\mbox{mb},
\eeq
as a range for different particles of type $i$, with the shorter time
applying to the early high-density stage.  
This is about an order of magnitude
shorter than the time scale for evolution of the fireball, which is
derived from the spatial size of the colliding system: for the largest 
nuclei, in particular the Pb--Pb or Au--Au collisions, 
over a wide range of energy, we expect
\beql{tauexest}
 \tau^{\mathrm{exp}}\simeq {R_A\over c}\simeq 5\mbox{--}8\,{\mbox{fm}/ c}.
\eeq

The achievement of kinetic equilibrium must be  visible in the thermal
energy spectra, and this behavior, as we argued above, can be understood
in qualitative terms using kinetic scattering theory for the case of 
nuclear collisions. However, it remains to date a mystery 
why in some important aspects thermal models succeed 
for the case of $p$--$p$
reactions. In particular, the exponential fall off
of particle spectra, suggesting thermal equilibrium, 
has been noted with trepidation for a considerable time. 

Hagedorn evaluated this behavior in the 
experimental data some 35 years ago \bcite{Hag65,Hag68,Hag73} and he
developed the statistical bootstrap model 
which  assumes a statistical phase-space distribution.
Hagedorn  called it {\it preestablished or
preformed equilibrium}: particles are produced in an elementary interaction
with a probability characterized by a universal temperature.
We can today only  speculate about the physical mechanisms.

In strong interaction physics it is possible that
vacuum fluctuations interfere with particle production
processes and could generate a preestablished 
thermal equilibrium distribution. In this case  very little
additional rescattering is needed for the development of 
thermal equilibrium. A possible mechanism is that the
production of quark pairs by snapping strings is subject to a
stochastic vacuum fluctuation force \cite{Bia99}, which results in 
natural Boltzmann distribution of produced quark pairs
at just the Hagedorn temperature, $T_{\rm H}=160$ MeV.
Another informally 
discussed possibility is the presence of intrinsic chaotic
dynamics capable of rapidly  establishing
kinetic equilibrium.

Sometimes, the fact that we do not fully understand 
thermalization in the $p$--$p$ case is 
raised as an argument against the possibility of 
conventional equilibration  in  nuclear collisions. 
We do not think so. In fact, if  the  $p$--$p$  case leads to 
thermal hadrons, we should have a yet better thermalization
in the $A$--$A$ case. Thus, a microscopic model that is adopted to extrapolate
from  $p$--$p$ to $A$--$A$ collisions should incorporate the concept of
the hadronic preestablished thermal equilibrium, else it is not going to be
fully successful.

The following stages are today believed to occur in
heavy-ion collision:
\begin{enumerate}
\item The initial quantum stage.\\
The formation of a thermalized state within 
$\tau_{\mathrm{th}}$
is most difficult to understand, and is subject to intense current
theoretical investigation.
During the pre-thermal  time, $0\le t<\tau_{\mathrm{th}}$, the properties of 
the collision system require the study both of 
quantum  transport and of decoherence phenomena, 
a subject reaching far beyond the scope of this article.
In this discussion, we assume that the thermal shape of  a
(quark, gluon) particle-momentum distribution is 
reached  instantaneously compared with the time scales for  chemical 
equilibration, see section \ref{Chem}.  This 
allows us to sidestep questions regarding the dynamics occurring 
in the first moments of the heavy-ion interactions, and we 
explore  primarily what happens after a time
$\tau_0\equiv \tau_{\mathrm{th}}\simeq 0.25$--$1$ fm/$c$.
The  value of $\tau_0$ decreases as the density of the pre-thermal 
initial state increases, \eg, as the  collision energy increases.
\item The subsequent  chemical equilibration time.\\ 
During the inter-penetration of the projectile and the
target lasting no less than $\sim$1.5 fm/$c$, diverse particle-production
reactions occur, allowing the approach to  chemical equilibrium
by light non-strange quarks $ q=u,\,d$. As the energy is redistributed 
among an increasing number of accessed degrees of freedom, the temperature 
drops rapidly.
\item  The strangeness chemical equilibration.\\
A third time period, lasting up to $\simeq$5 fm/$c$, during which the
production and chemical equilibration of strange quarks takes place.
There is a reduction of temperature now mainly due to the expansion flow, 
though the excitation of the strange quark degree of freedom also
introduces a non-negligible cooling effect. 
\item The hadronization/freeze-out.\\
The fireball of dense matter expands and decomposes into
the final state hadrons, possibly in an (explosive)
process that does not allow  re-equilibration of 
the final-state particles. The dynamics is 
strongly dependent on the size of the initial state
and on the nature of the equations of state.  
\end{enumerate}
Throughout these stages, a local thermal equilibrium is rapidly 
established and, as noted,  the local temperature evolves in time 
to accommodate change in the internal structure and size,
 as is appropriate for
an isolated physical system. We have a temperature
that passes through the following nearly separable series of 
stages:
\begin{center}
\begin{tabular}{cl}
$T_{\mathrm{th}}$&\hspace*{-0.3cm} the temperature associated with the 
initial thermal equilibrium,\\

$\downarrow$&\hspace*{0.cm}
{\it evolution dominated mainly by production of\/} $ q$ and ${\bar q}$;\\

$T_{\mathrm{ch}}$&\hspace*{-0.3cm} chemical equilibrium of non-strange 
quarks and gluons,\\

$\downarrow$&\hspace*{0.cm}
{\it evolution dominated by  expansion and 
production of\/} $ s$ {\it and} ${ \bar s}$; \\

$T_{\mathrm s}\ $&\hspace*{-0.3cm}
condition of  chemical equilibrium of $ u,d$ and $s$ quark flavors,\\

$\downarrow$&\hspace*{0.cm}
{\it expansion, dissociation by particle radiation\/};\\

$T_{\mathrm f}$&\hspace*{-0.3cm}
temperature at hadron-abundance freeze-out, \\

$\downarrow$&\hspace*{0.cm}
{\it hadron rescattering, reequilibration\/}; and\\

$T_\mathrm{tf}$&\hspace*{-0.3cm}
temperature at thermal freeze-out, $T_\mathrm{tf}=T(\tau^{\mathrm{exp}})$.
\end{tabular}
\end{center}
\vspace{-0.1cm}

\noindent We encounter a considerable decrease
in temperature.  The entropy content of an evolving 
isolated system must increase, and this is initially 
related to the increase in the number of particles within the 
fireball and later also due to the
increase in volume. However,  in the later stages dominated by
flow, the practical absence of viscosities in the quark--gluon 
fluid implies that 
there is little additional  production of entropy. 
The final entropy content is
close to the entropy content established in the  earliest thermal stage of the
collision at $t<\tau_0$, despite a drop in temperature by as much as a factor 
of two under current experimental RHIC conditions during the
sequel  evolution of the fireball. 

Except for the unlikely scenario of a fireball not expanding, 
but suddenly disintegrating in the early stage,
none of the temperatures discussed above corresponds to
the temperature one would read off the (inverse) 
slopes of  particle  spectra. In principle, the 
freeze-out temperature determines the shape of emission and 
multiplicity of emitted particles. However, the freeze-out occurs 
within a local flow field of expanding matter and the thermal
spectrum is to be folded with the flow which imposes
a Doppler-like shift of $T_{\mathrm{tf}}$: we observe a higher temperature 
than is actually locally present when particles decouple from flowing
matter (kinetic or thermal freeze-out). The observable  temperature $T_\bot$ 
is related to the intrinsic  temperature   of the source: 
\beql{doppler}\label{eq:dopplers}
T_{\bot}
\simeq
{ {1+\vec n \cdot  \vec v_\mathrm{tf}}\over 
     {\sqrt{1-\vec v^{\,2}_\mathrm{tf}}}} T_\mathrm{tf}
\to
\sqrt{1+v_\mathrm{tf}\over 1-v_\mathrm{tf}} T_\mathrm{tf}.
\eeq
This relation must be used with caution, since it does not apply
in the same fashion to all particles and has a precision rarely better than
$\pm$10\%. 

\begin{figure}[tb]
\vspace*{-1.1cm}
\centerline{\hskip 0.5cm
\epsfig{width=9.cm,clip=,angle=0,figure=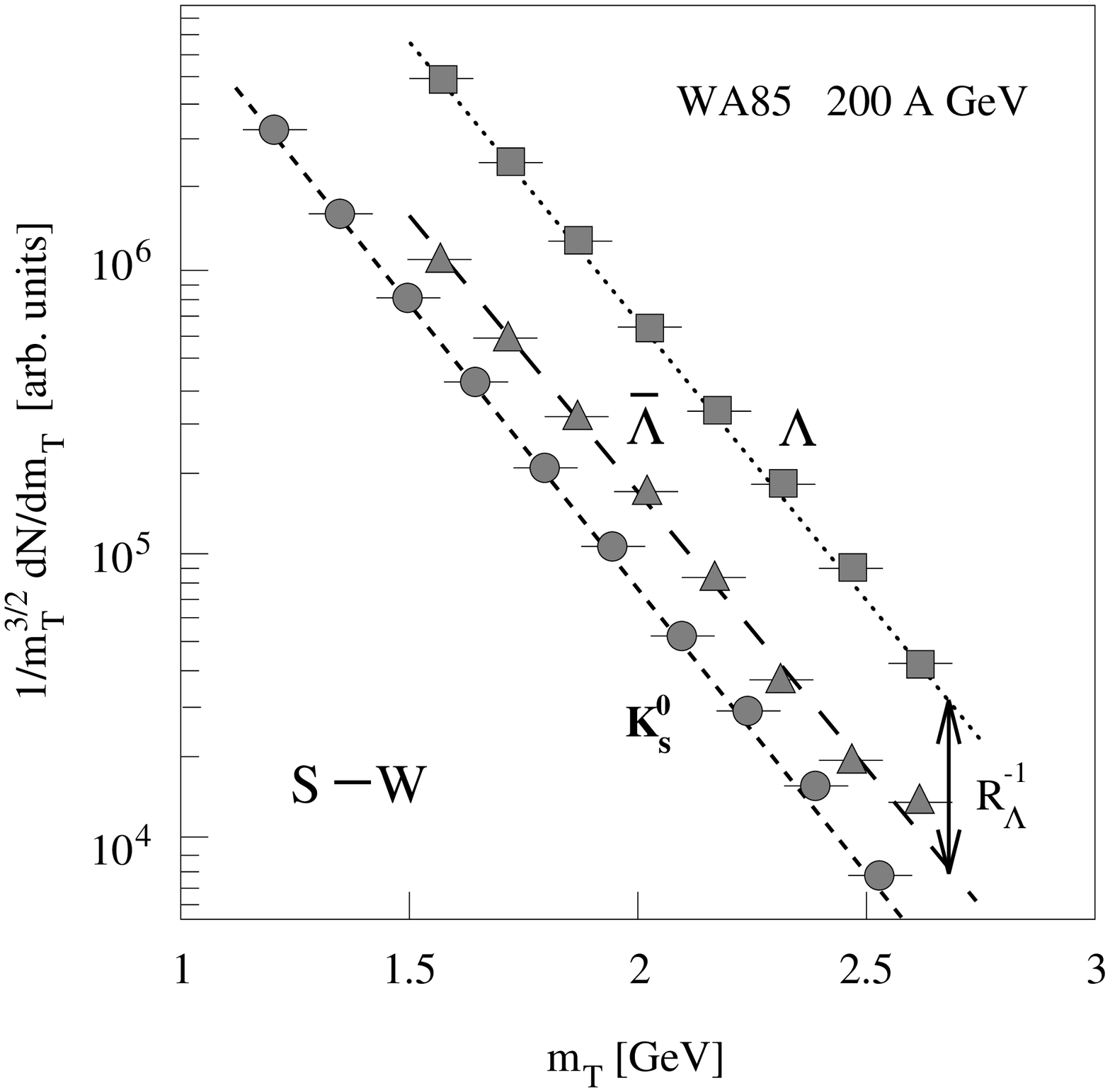}
}
\vspace*{-0.4cm}
\caption{ 
Strange particle spectra for $\Lambda,\ \overline{\Lambda},$ and 
${K}_{\rm S}$ \cite{Raf96a}. The line
connecting the $\Lambda$ and $\overline{\Lambda}$ spectra, denoted 
$R_\Lambda^{-1}$, shows how at fixed $m_\bot$ the ratio $R_\Lambda$ of 
abundances of these particles can be extracted. 
Experimental WA85 results at 200\agev
\protect \cite{Eva94,Eva95,Dib95}.
\label{specWA85}
}
\end{figure}

To check if thermalization (momentum equilibration) is established, we  
consider momentum
distributions in the direction transverse to the collision axis.
Under a Lorentz transformation along the collision axis, $p_\bot$ remains 
unchanged and thus
\[
m_\bot=\sqrt{m^2+\vec p_\bot^{\,2}},
\]
is invariant. Transverse mass $m_\bot$-particle spectra are not directly
distorted by  flow  motion  of the fireball matter along 
the collision axis, and also no further consideration of 
the CM frame of reference is necessary, which in fixed target experiments 
is rapidly moving with respect to a laboratory observer. 

In order to study the thermal properties in the fireball 
as `reported' by the emitted particles, we analyze
 $m_\bot$ spectra of many different hadrons. The 
range of $m_\bot$,  on the one hand, should not reach very large values, 
at which  hadrons originating in hard parton scattering
are relevant. On the other hand, we do study relatively small 
$m_\bot$, in order to avoid  the  non-exponential structure associated 
with  transverse matter flow and unstable resonance decays.

The  central-rapidity  high-transverse-mass spectra
of strange particles, K$_{ s}^0$, 
$\overline\Lambda$, and $\Lambda$, given by the 
CERN--SPS WA85 collaboration  \cite{Eva94,Eva95,Dib95}, 
$m_{\bot}^{-3/2}\,dN_i/dm_{\bot}$, 
are shown in figure \ref{specWA85} on a semi-logarithmic display. The 
spectra can be  fitted with a straight line. Similar results
were also reported from the related work of the WA94 collaboration 
 for S--S interactions \bcite{Aba95b}. 
We  see, in figure \ref{specWA85}, in the
region of transverse masses presented,  1.5 GeV $<m_{\bot}<2.6$ GeV,
that the behaviors of all three  different particles feature the same 
inverse exponential slope,  $T_{\bot}=232\pm5$ MeV. This is not the actual 
temperature of the fireball, as noted in \req{dopplers}. 

\begin{figure}[tb]
\centerline{\hskip 0.5cm
\epsfig{width=14.cm,clip=,angle=0,figure=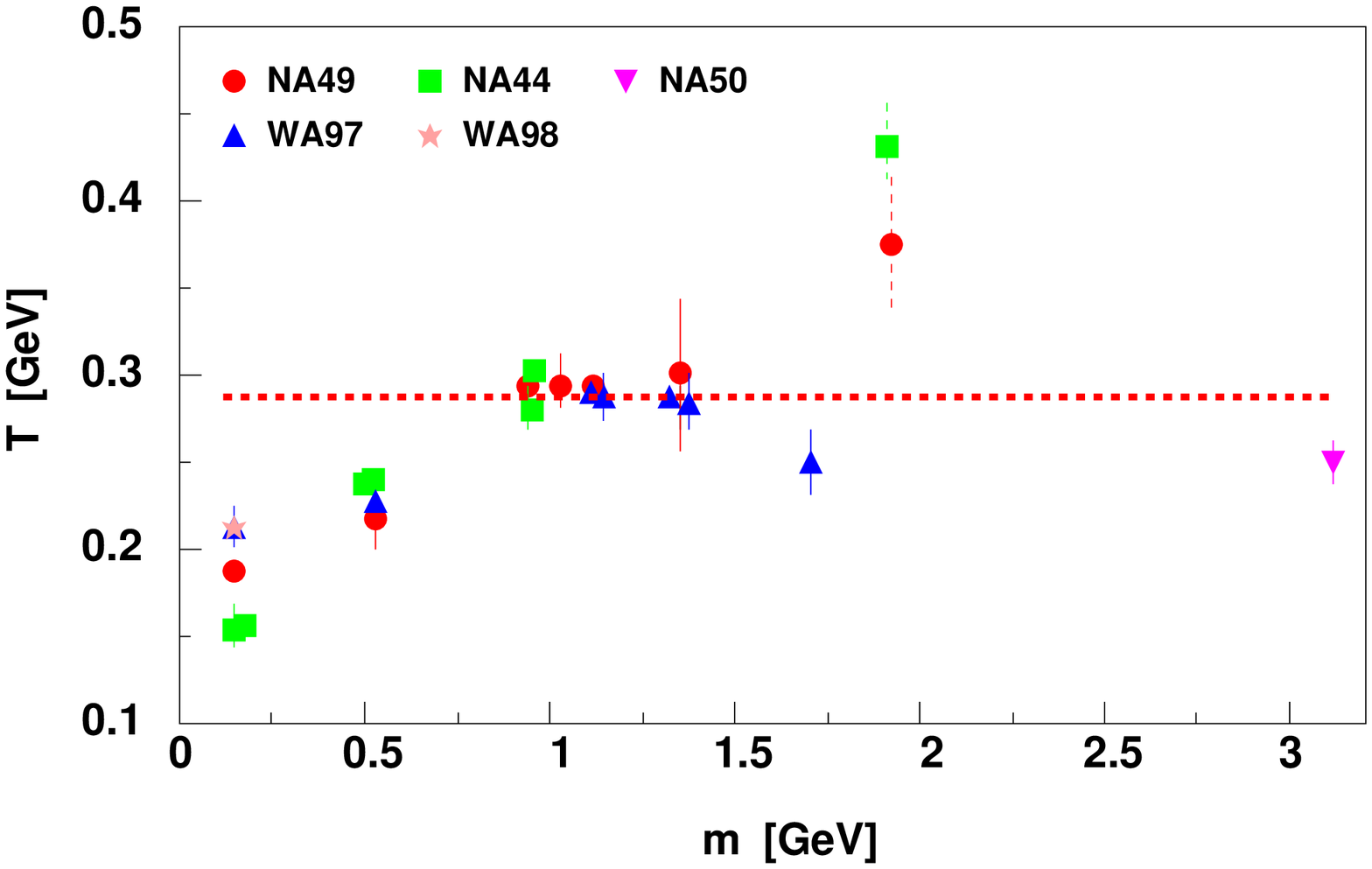}
}
\caption{ 
Inverse slopes $T_\bot$ observed in Pb--Pb 
interactions at 158\agev as function of particle mass; 
symbols indicate the experiment from
which data is drawn, as coded in the figure.
\label{PbPbtemp}
}
\end{figure}

Presence of matter flow which generates the relatively high value
of spectral slope is also responsible for differences in the 
observed values of $T_\bot$ for particles of widely different 
mass such as pions, kaons and nucleons considered in a similar
$p_\bot$ interval. This is illustrated in 
figure \ref{PbPbtemp}. A remarkable feature is that many
strange baryons are seen to have similar inverse slope, as 
is also seen in table \ref{WA97temp}.

\begin{table}[tb]
\caption{\label{WA97temp}
Inverse  slopes $T_\bot$ 
for various strange hadrons.  }
\begin{tabular}{ccccccc}
\hline\hline
$\!$$\vphantom{\ds\frac{e}{.}}$ &$\!$$\Lambda$&$\!$$\overline{\Lambda}$&$\!$$\Xi^-$
&$\!$$\overline{\Xi}^+$&$\!$$\Omega^-+\overline{\Omega}^+$&$\!$${\phi}$\\  
\hline
$\!$Pb--Pb          &$\!$$289\pm2$  &$\!$$287\pm4$ &$\!$$286\pm9\phantom{2}$  
&$\!$$284\pm17$ &$\!$$251\pm19$  &$\!$$305\pm15$   \\
$\!$S--W          &$\!$$233\pm3$  &$\!$$232\pm7$ &$\!$$244\pm12$  &$\!$$238\pm16$ 
&$\!$            &$\!$                 \\
\hline
\hline
\end{tabular}
\end{table}

\section{Chemical equilibria}\label{Chem}
The average energy of each particle defines 
the local temperature $T$, good for all particles.
A local chemical potentials $\sigma_i$ need to be introduced
for each kind of particle `$i$' in order to establish the particle 
 density. We will see that it is often more convenient
to use  particle fugacity,
\begin{equation}\label{Upsilon}
\Upsilon_i\equiv e^{\sigma_i/T}.
\end{equation}
The statistical parameters $T$ and $\sigma_i$ (or $\Upsilon_i$) 
express different types of equilibration 
processes in the hadron matter fireball, and in general
there is a considerable difference in the time needed
for the attainment of
thermal  and, respectively,  chemical equilibrium.

The fugacity controls, independently of temperature, the
yield of the particle species.  For Boltzmann statistics, 
the fugacity $\Upsilon_i$ 
is multiplying the particle distribution, which in local rest frame
assumes the usual form,
\begin{equation}\label{BolDis}
{{d^6N_i}\over{d^3pd^3x}}=g_i{\Upsilon_i\over (2\pi)^3}e^{-E_i/T}.
\end{equation}
For the Fermi/Bose quantum distributions the fugacity factor remains in 
front of the exponential, 
\begin{equation}\label{FDDis}
{{d^6N_i^{\rm F/B}}\over{d^3pd^3x}}= {g_i\over (2\pi)^3}
{1\over \Upsilon_i^{-1}e^{E_i/T}\pm 1},
\end{equation}
The coefficient $g_i$ is the degeneracy of the particle considered, 
it comprises the  intrinsic properties such as spin
and color of the particle. 
 
The number of particles `$i$' present is the result of momentum integral of 
the  phase-space distribution. Thus 
the yield of particles depends on both temperature and 
the fugacity. However, fugacity  practically does not control
the shape of the momentum distribution, and thus it is generally
said that it can be chosen to produce the required 
particle density, with temperature being the parameters determining
the momentum distribution.  

The allowed range of the (always positive) 
fugacities is not restrained for Boltzmann and Fermi particles. However
for bosons, the condensation singularity of the distribution cannot 
be crossed, which  requires that 
\begin{equation}\label{UpsilonLim}
\Upsilon_i\le e^{m_i/T}.
\end{equation}
As usual the $E_i(p)>m_i$ for all $p$ and  $m_i$ is the
particle rest mass. Almost never the exceptional 
value $\Upsilon_i=1$ applies, with the exception of 
an chemically equilibrated gas of particles (such as photons)
which do not carry any charge, \ie, a conserved, discrete quantum number. 

As the number of particles evolves in a reaction, the fugacities 
usually change. For example, we begin with small initial  yield of strange
quarks and antiquarks in a deconfined quark-gluon plasma. Subsequent reactions
of the type,
\begin{equation}\label{sprodmech}
GG\to s\bar s,\qquad   q \bar q\to  s\bar s,
\end{equation}
increase the yield of pairs of strange quarks and if we could
cook the hadronic matter at a constant temperature $T$ (thermal bath),
than after
some time known as the chemical relaxation time, the yield of
strange quarks and antiquarks would approach the chemical equilibrium
yields corresponding to the fugacity approaching unity,
\begin{equation}\label{Upsilon1}
\Upsilon_{i=s}(t)=\Upsilon_{i=\bar s}(t)\to 1\,.
\end{equation}
The important message here is that the particle
fugacities are time dependent and can evolve rapidly during the 
heavy-ion collision process. Description of the dynamical evolution
of the fugacity is one of the challenges we are facing in order to
understand the physics of hadronization.  

Another important challenge which is arising, is due to the presence of  
two quite different types of chemical equilibria.
\begin{itemize}
\item In relativistic reactions, particles can be made
as energy is converted to  matter. Therefore, we can expect 
to approach  slowly the {\it absolute} chemical 
equilibrium. Absolute chemical equilibrium is hard
to grasp intuitively, since our instincts are 
distorted by the fact that a black body photon radiator 
is correctly assumed to be in absolute chemical equilibrium 
on the time scale defined by a blink of an eye.
We characterize the approach to absolute 
chemical equilibrium by a  fugacity factor 
$\gamma_i$ for particle `$i$'. We will evaluate the 
evolution of $\gamma_i$ in the heavy-ion 
collision reaction as a function of time.
Given its physical meaning it is often referred to as the
phase-space occupancy. 
\item {\it Relative} chemical 
equilibration, the case
commonly known   in chemistry. It involves reactions that 
distribute or maintain a certain already existent element among 
different accessible compounds. Use of chemical  potentials 
associated with conserved global properties, such as $\mu_b$
for baryon number,  presupposes that the particular relative chemical 
equilibrium is present. For example production of quarks in pairs 
assures that there is always the same net flavor (thus baryon number) 
balance, and hence the relative
flavor equilibrium is maintained. The chemical
potential of antiparticles is in consequence 
the negative of that for particles, {\it provided} that we 
have introduced appropriate $\gamma_i$ to control the absolute
yield of particle-antiparticle pairs.
\end{itemize}

In general the fugacity of each individual particle will comprise the 
two chemical factors associated with the two different chemical equilibria. 
For example, let us look at the nucleon, and the antinucleon: 
\begin{equation}\label{UpsilonN}
\Upsilon_N=\gamma_N e^{\mu_b/T},\qquad  
\Upsilon_{\overline{N}}=\gamma_N e^{-\mu_b/T}.
\end{equation}
Equivalently, we can keep  separate 
 chemical potentials for particles and antiparticles \bcite{Mat86,Mat86b},
\begin{equation}\label{sigmaN}
\sigma_{N}\equiv \mu_b+T\ln\gamma_N,  \qquad
\sigma_{\overline{N}}\equiv -\mu_b+T\ln\gamma_N.
\end{equation}

There is an obvious difference between the two chemical 
factors in Eq.\,(\ref{UpsilonN}):
the number of nucleon-antinucleon pairs is associated
with the value of $\gamma_N$ but not with\,$\mu_b$. This can be seen looking at the first 
law of thermodynamics, in this context written as:
\beqarl{1Emu}\nonumber
dE&=&-P\,dV+T\,dS+\sigma_N\,dN+\sigma_{\overline{N}}\,d\overline{N}\\
&=&
-P\,dV+T\,dS+\mu_b(dN-d\overline{N})+T\ln\gamma_N(dN+d\overline{N}).
\eeqar
To obtain the second form we have employed Eq. (\ref{sigmaN}). 
We see that $\mu_b$ is the energy required to change 
the baryon number, 
$$b\equiv N-\overline{N},$$
 by one unit, while the
number of nucleon-antinucleon pairs, 
$$2N_{\rm pair}\equiv  N+\overline{N},$$
is related to $\gamma_N$. For $\gamma_N=1$ the last term vanishes, at this
point small fluctuation in number of nucleon pairs does not influence the 
energy of the system, we are have reached the  absolute baryochemical 
equilibrium.

Presence of $\gamma_N$  allows us to count and control within a theoretical
description how many pairs of nucleons are added. When $\gamma_N\to 1$,
we have as many as we would expect in absolute chemical equilibrium,
of course establishing chemical equilibrium for anti- nucleons
 can take a long time. However, it  is  not  uncommon to see in contemporary  
literature assumption of instantaneous absolute  
chemical  equilibrium,  $\gamma_i\to 1$,
with  particles `instantaneously' reaching 
their  absolute  chemical equilibrium 
abundances.  The argument presented is that such a simple model `works',
as it is capable to describe widely different particle yields qualitatively.
This in fact is the observation which motivated Hagedorn's work. 40 years
after, our foremost interest is to understand the deviations 
from this, in order to unravel the mechanism by which such a 
surprising result can arise. 

\begin{figure}[tb]
\vspace*{-1.1cm}
\centerline{\hskip 0.5cm
\epsfig{width=4.cm,clip=,angle=-90,figure=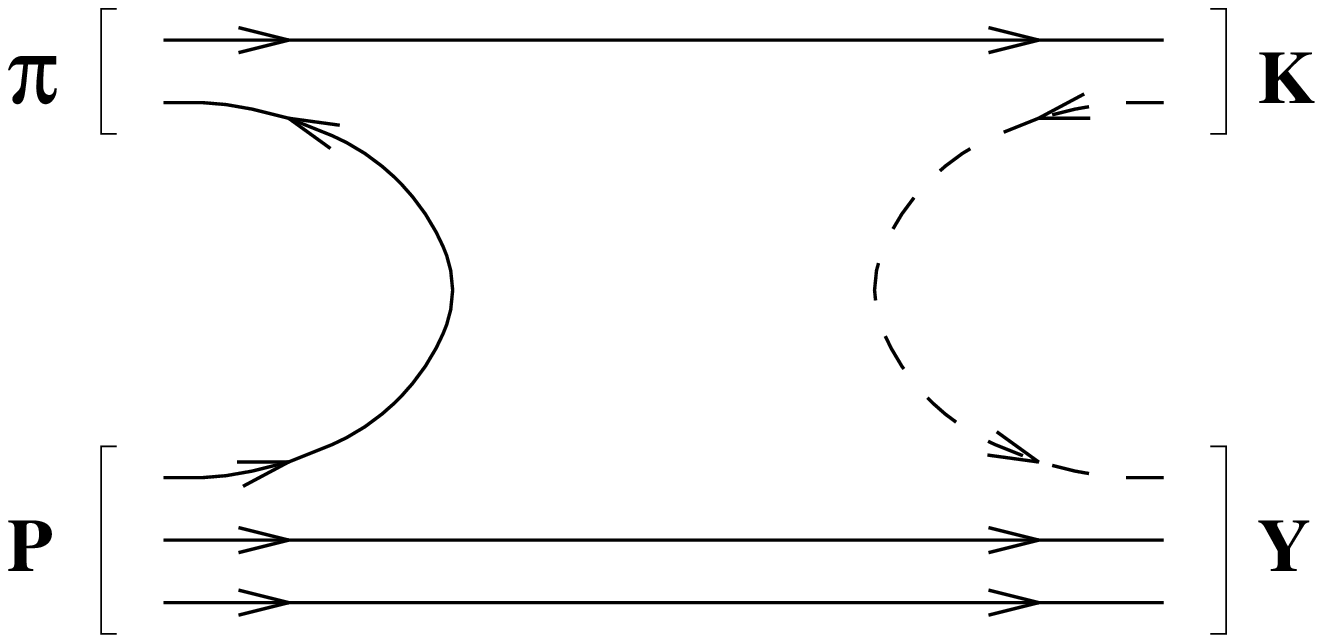}
}
\vspace*{-0.4cm}
\caption{ 
The production of strangeness
in reactions of the type  ${\pi}+N\to K+Y$ in the 
HG phase. Solid lines indicate
the flow of light quarks and the disappearance 
of one $\bar qq$ pair, the dashed line
is for the added $\bar s s$ pair.
\label{sprodHG}
}
\end{figure}

In order to better understand the difference between absolute and 
relative chemical equilibrium, let us consider
 the abundance of strangeness in the baryon-rich 
hadronic gas (HG) phase. There is little `strangeness'   initially, 
strangeness is made in collision processes of the interacting 
hadronic matter. 
We begin far from absolute  chemical-strangeness equilibrium. 
To make $ s\bar s$ pairs in a HG  phase, there
are many possible reactions, classified usually as the 
direct- and associate-production 
processes. In the associate production process, a pair of strange quarks 
is shared between two existent hadrons, of which one is a baryon, typically
a nucleon N, which becomes a hyperon Y:
\[
{\pi}+\mathrm N\leftrightarrow \mbox{K}+\mbox{Y}.
\]
This situation is illustrated in figure \ref{sprodHG}.

In a direct-production process, a  pair of strangeness-carrying particles is formed 
directly via annihilation of two mesons,  
adding a pair to the system:
\[
{\pi}+{\pi}\leftrightarrow \mbox{K}+\overline{\mbox{K}}.
\]
Here, a   pair of strange particles is made in the form of a 
pair of kaons, K$^+$K$^-$.

\begin{figure}[tb]
\vspace*{-1.1cm}
\centerline{\hskip 0.5cm
\epsfig{width=4.cm,clip=,angle=-90,figure=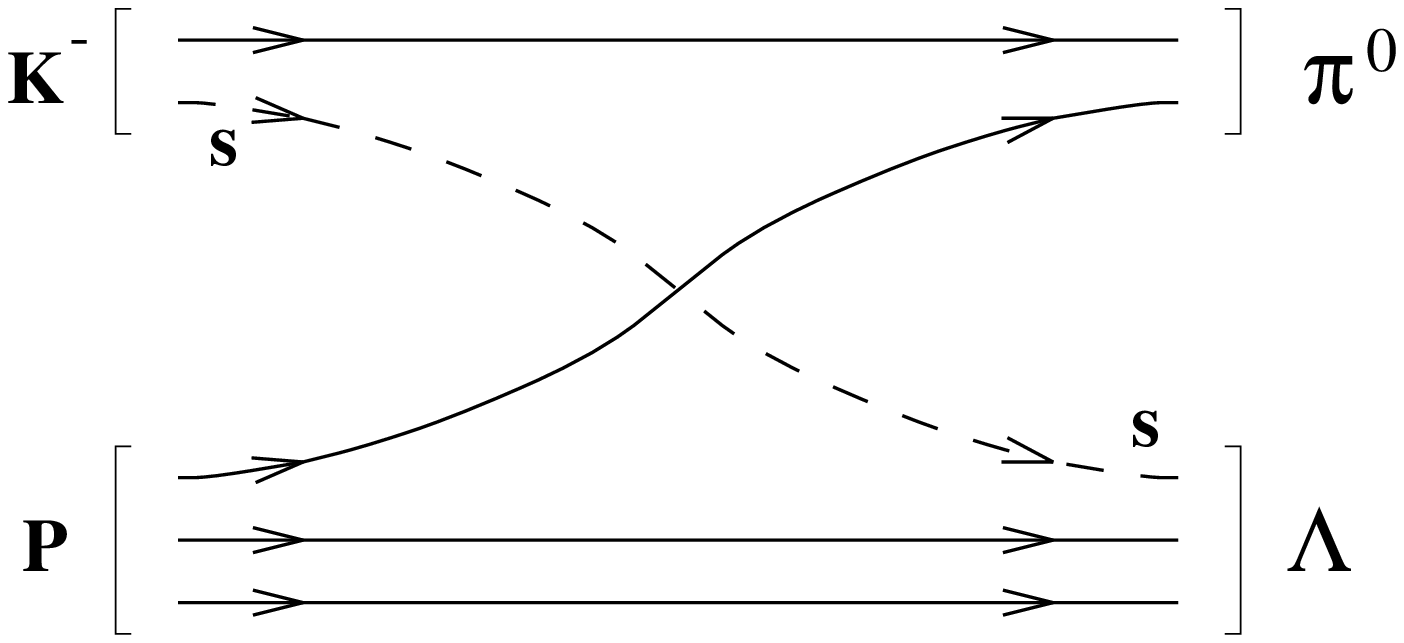}
}
\vspace*{-0.4cm}
\caption{ 
An example of a strangeness-exchange reaction in the HG phase:
$ K^-+p\to \Lambda+{\pi}^0$. Solid lines,
flow of $ u$ and $d$ quarks; dashed line, exchange of an 
$ s$ quark between two hadrons.
\label{strexch}
}
\end{figure}

With these two reaction types alone it could be that populations of strange 
mesons and baryons evolve differently. However, 
the meson carrier of the $s$ quark, K$^-$, can exchange this quark, 
see figure \ref{strexch}, via fast  exothermic reaction with a nucleon, 
forming a hyperon:
\[
\mbox{K}^-+\mathrm N\leftrightarrow {\pi}+\mbox{Y}.
\]
This reaction establishes relative chemical equilibrium between mesons and
baryons by 
being able to move the strange quark between these two different 
strangeness carriers, $ s\bar q$ mesons  and $ sqq$ baryons.

Reactions establishing the 
redistribution of existent flavor, or the abundance of some other 
conserved quantity, play a different role from the 
reactions that actually contribute to the 
formation of this flavor, or other quantum number, 
and facilitate the approach to absolute 
chemical equilibrium. Accordingly, the  
time constants for relaxation are different, 
since different types of reaction are involved. 

Apart from the different relaxation times associated with the different 
types of thermal and chemical equilibria, there are different time 
scales associated with the different fundamental interactions
involved. For example, the electro-magnetic interactions are
considerably slower at reaching equilibrium than are the strong interactions 
governing the evolution of dense hadronic fireballs created in  
ultra-relativistic heavy-ion collisions. 
All the important time constants for relaxation in heavy-ion collisions
arise from  differences in mechanisms operating within the realm 
of strong interactions. Therefore,
the separation of time scales is not as sharp as that between
the different interactions, though a clear hierarchy arises and we
have presented it above.

There are just a few chemical potentials which suffice
in the study of hadronic matter.  It is convenient to 
use the quark flavor for this purpose, since there is a continuity 
in the notation across the phase boundary of quark matter and hadronic
matter. We thus use:
\begin{equation}\label{uds}
\lambda_u=e^{\mu_u/T},\quad \lambda_d=e^{\mu_d/T},\quad \lambda_s=e^{\mu_s/T} . 
\end{equation}
Since more often than not the light flavors $u,d$ remain indistinguishable
one often uses:
\begin{equation}\label{udq}
\mu_q=\frac12 (\mu_u+\mu_d),\quad \lambda_q^2=\lambda_u\lambda_u.
\end{equation}
Considering that three quarks make a baryon we also have
\begin{equation}\label{q3}
\mu_b=3\mu_q,\quad  \lambda_b=\lambda_q^3.
\end{equation}
It is important to remember that the three quark flavors $u,d,s$ carry baryon 
number. Thus when baryon density is evaluated this has to be
appropriately allowed for, counting all three quark densities. 
This can be easily forgotten for strangeness, considering the usual 
definition of baryochemical potential, Eq. (\ref{q3}).

At this point, we recall the relationship of strange quark chemical potential
and strangeness chemical  $\mu_{\rm S}$
potential (as opposed to strange quark chemical 
potential  $\mu_s$):
\begin{equation}\label{Ssrel}
\mu_s=\frac 1 3 \mu_b-\mu_{\rm S},\quad \lambda_s={\lambda_q\over\lambda_{\rm S}},
\end{equation}
which expresses the fact that strange quarks carry negative strangeness 
and one third of baryon number. Consequently, hyperons have negative 
strangeness, antihyperons positive strangeness, ${\rm K}^-(\bar us)$
negative strangeness and thus it is called antikaon $\overline{\rm K}$,
 and ${\rm K}^+(u\bar s)$ has positive strangeness, and it is the kaon K.
This nomenclature has an obvious historical origin, when strangeness was
first discovered, the $\bar s$ containing K$^+$ has been
 the particle produced.

Using Eq. (\ref{Ssrel}) in statistical formulation
allows to evaluate baryon number and strangeness by simply differentiating the 
grand canonical partition function $\cal Z$: 
\begin{equation}\label{ZS}
N_b= \lambda_b
{\partial{\cal Z}(\beta,\lambda_b,\lambda_{\rm S})\over \partial\lambda_b},\quad 
N_{\rm S}=  \lambda_{\rm S}{\partial{\cal Z}(\beta,\lambda_b,\lambda_{\rm S})\over 
\partial\lambda_{\rm S}},\quad 
\end{equation}
where as usual $\beta=1/T$. Especially the computation of $N_b$ is more
cumbersome when $\lambda_i, i=u,d,s$, is used.

On the other hand, we believe that the use of fugacities which 
follow the yield of quarks, $\lambda_i, i=u,d,s$,
is intuitive and easy in programing, mistakes are rather difficult to make.
Particle yields can be easily followed and checked, and thus 
errors and omissions minimized.
As an example, let us consider the ratio of a baryon and antibaryon with
strangeness and for reasons which will become obvious in a moment 
we choose to look at the ratio $\overline{\Xi^-}(\bar d \bar s \bar s)/\Xi^-(dss)$.
Given the quark content and ignoring isospin asymmetry we find:
\begin{equation}\label{Xirat}
{\overline{\Xi^-}\over \Xi^-}=
{\lambda_s^{-2}\lambda_q^{-1}\over\lambda_s^{2}\lambda_q}
=\lambda_s^{-4}\lambda_q^{-2}\quad =
{e^{2\mu_{\rm S}/T}e^{-\mu_b/T}\over e^{-2\mu_{\rm S}/T}e^{\mu_b/T}}
=e^{4\,\mu_{\rm S}/T}e^{-2\mu_b/T}.
\end{equation}
Since all $\Xi$ resonances which contribute to this ratio are symmetric
for particles and antiparticles, and possible weak interaction 
feed from $\overline\Omega(\bar s \bar s \bar s) $ and, respectively
 $\Omega(sss)$ are small, these expressions are actually rather exactly 
giving the expected experimental ratio. 

When we check the results presented by \cite{Bra01}, we find for
the freeze-out parameters stated there $T=174$ MeV, 
$\mu_b=46$ MeV and $\mu_{\rm S}=13.6$ MeV, 
a ratio $\overline{\Xi^-}/\Xi^-=0.806$, which differs from the result
$\overline{\Xi^-}/\Xi^-= 0.894$ given. 
While the difference appears small, when looked at in terms of the 
particle chemical potential this is a big effect: 
Considering that 
the Cascades have baryon number $b$ = 1 and strangeness S $=-2$, we expect
that the $\Xi$-chemical potential is $
\mu_\Xi|_{\rm expected}=\mu_b-2\mu_{\rm S}=18.8\, \mbox{MeV}$.
On the other hand given the result seen in   \cite{Bra01},
\[
0.894={\overline{\Xi^-}\over \Xi^-}=e^{-2\mu_\Xi/T}.
\]
The particle chemical potential is $
\mu_\Xi|_{\rm used}=9.75\, \mbox{MeV}$.
In order to approach using the stated potentials
the answer presented in \cite{Bra01},
we can try to use another quark fugacity definition,
\begin{equation}\label{XiratPBM}
\left.{\overline{\Xi^-}\over \Xi^-}\right\vert_{\rm PBM}\equiv 
{\lambda_s^{2}\lambda_q^{-1}\over\lambda_s^{-2}\lambda_q}
 =e^{\frac 23\mu_b/T-4\mu_{\rm S}/T}=0.87.
\end{equation}
However, this redefinition is not consistent with the ratios of particles
such as $K^-/K^+$ also seen in \cite{Bra01}.
This example of a possible partial misunderstanding of the quantum
numbers associated with different hadrons illustrates
the danger of using anything but the (valance) quark
fugacities in study of chemical properties of hadrons.

To complete the discussion of chemical potentials, we need to address
the $u$--$d$ asymmetry which is only noticeable  for Pb--Pb interactions
at SPS. We complement Eq. (\ref{udq}) by introducing 
\begin{equation}\label{delmu}
\mu_I=\mu_d-\mu_u,\quad  \lambda_I=\frac{\lambda_d}{\lambda_u}\ge 1.
\end{equation}
In the last inequality, we show the constraint arising 
 given that nuclei have in general a greater neutron than proton number.
The subscript $I$ reminds us of isospin, we have also used $\delta\mu$ for 
$\mu_I$ in the past. A sensitive  probe of $\lambda_I$ is the ratio
\begin{equation}\label{delmupi}
{\pi^-(\bar u d)\over \pi^+(u \bar d)} =\lambda_I^2.
\end{equation}

At RHIC, for $\sqrt{s_{NN}}=130$ GeV, 
the outflow of projectile and target matter from 
the central rapidity region leaves a very small `input' isospin asymmetry:
with a baryon density per unit rapidity at $dN_b/dy\simeq 20$, but hundreds
of mesons made, the valance quark asymmetry is negligible. 
However, at the top energy at SPS there is an effect at the level of 2--4\%
percent, since baryon number is retained in the fireball and 
the meson to baryon ratio is 8 times smaller, as measured in 
terms of the ratio of all negative hadrons,
\begin{equation}\label{hm}
h^-=\pi^-+K^-+\bar p,
 \end{equation}
to baryon number. At SPS, we have $h^-/b\simeq 2$, while at RHIC at 
$\sqrt{s_{NN}}=130$ GeV,  $h^-/b\simeq 16.5$, in both cases at central rapidity.
We can estimate for SPS where  $\lambda_q=1.6$:
 $$1.6<\lambda_d<1.66,\quad  1.54<\lambda_u<1.6\,.$$

The exact value of $\lambda_d, \lambda_u$ depends on what form of matter
(confined, deconfined) is the source of particles produced, and the influence
of gluon fragmentation into quark pairs. The limiting values are found
ignoring gluon fragmentation and evaluating the ratio of of all
down and up quarks in hot quark matter, where quark 
density is $\rho_i\propto \mu_iT^2, i=u,d$. Thus we find 
\begin{equation}\label{mdmu}
\frac{\rho_d}{\rho_u}=1+\delta=\frac{\mu_d}{\mu_u},
\quad\delta=\frac{n-p}{n+2p},
\end{equation}
where $ n(udd)$ and $p(uud)$ are the neutron, and respectively, proton input
into the quark matter source.  For heavy nuclei $\delta\simeq 0.156$, and
we find:
\begin{equation}\label{luld}
\lambda_d=\lambda_q^{\frac 1 {1-\delta/2}},\quad 
\lambda_u=\lambda_q^{\frac 1 {1+\delta/2}}.
\end{equation}

We now return to discuss further the characterization  of chemical non-equilibrium.
When we study the approach to chemical equilibrium for
different hadrons, we can use  three non-equilibrium parameters,
$\gamma_u,\gamma_d,\gamma_s$, and equivalently, $\gamma_q^2=\gamma_u\gamma_d$. 
In quark matter, these three factors express the approach to 
the expected chemical equilibrium yield by the quark abundances. Upon
hadronization quarks are redistributes among all  individual hadrons
and the non-equilibrium abundances can be characterized by the same three
factors only, since in all hadron formation reactions only quark-antiquark 
pairs of the same flavor can be formed. 
It is important to realize that even if there were no
change of the quark pair number in hadronization, 
the values of $\gamma_u,\gamma_d,\gamma_s$
in hadron gas and quark matter must differ since the phase 
spaces have different size as we shall discuss further below.
In general, we have to distinguish   
$\gamma_u^{\rm QGP},\gamma_d^{\rm QGP},\gamma_s^{\rm QGP}$
from  $\gamma_u^{\rm HG},\gamma_d^{\rm HG},\gamma_s^{\rm HG}$.
Moreover, as noted these sets of parameters differ due to hadronization of
 gluons into quark pairs.

We thus characterize the fugacities of all hadrons by six parameters.
For baryons the typical examples are (considering protons $p(uud)$,
antiprotons $\bar p(\bar u\bar u\bar d)$, $\Lambda(uds)$, 
$\overline\Omega(\bar s\bar s\bar s)$ as examples):
\begin{equation}\label{gambar}
\Upsilon_p=\gamma_u^2\gamma_d\,e^{2\mu_u+\mu_d},\ 
\Upsilon_{\bar p}=\gamma_u^2\gamma_d\,e^{-2\mu_u-\mu_d},\ 
\Upsilon_\Lambda=\gamma_u\gamma_d\gamma_s\,e^{\mu_u+\mu_d+\mu_s},\ 
\Upsilon_{\overline\Omega}=\gamma_s^3\,e^{-3\mu_s}.
\end{equation}
The yield of mesons follows the same  pattern (considering 
$\pi^+(u\bar d)$, $\pi^-(\bar u d)$, $K^-(\bar u s)$, 
$\phi(\bar s s)$ as examples):
\begin{equation}\label{gammes}
\Upsilon_{\pi^+}=\gamma_u\gamma_d\,e^{\mu_u-\mu_d},\quad
\Upsilon_{\pi^-}=\gamma_u\gamma_d\,e^{-\mu_u+\mu_d},\quad
\Upsilon_{K^-}=\gamma_u\gamma_s\,e^{-\mu_u+\mu_s},\quad
\Upsilon_{\phi}=\gamma_s^2.
\end{equation}

It is important to note that this approach assumes that the relative
population of heavier resonances is in chemical equilibrium with the
lighter states, since we have the same value of $\Upsilon$ for
all hadrons with the same valance quark  content. For example
$\Upsilon_p=\Upsilon_{\Delta^+}$. We thus realize that this approach 
solely focuses on the quark distribution  and does not allow for the 
possibility that heavier resonances may simply not be populated.
This method thus is most suitable for 
a hadronizing quark matter fireball, and may miss important 
features of a hadron fireball which never entered the deconfined phase. 
When conditions are present which let us believe that the 
population of important heavier resonance states such as $\Delta(qqq)$
could be suppressed beyond the thermal factor $\propto e^{m_\Delta/T}$ 
we have to restore a factor $\gamma_\Delta < \gamma_q^3$ which allows for 
such suppression. As long as this is not done, we in fact
assume that the relative abundances of hadronic states of same
quark content are in chemical equilibrium.  The following study
of deviations from chemical equilibrium is addressing 
a much more subtle question, namely how the overall number
of (valence) quarks compares to the hadron phase-space 
chemical equilibrium expectations. 

\section{How near to chemical equilibrium?}\lssec{chemeqQM} 
Is such a characterization of hadron abundances at all functioning, and
is so, how close are we actually to absolute chemical equilibrium in heavy-ion
reactions? To answer this question a systematic test was performed 
for the S--W/Pb 200\agev experimental SPS results \cite{Let99}. 
Particle yields
were fitted allowing progressively greater and greater degree of
chemical non-equilibrium. The results in table \ref{fitsw}
show that the statistical significance is increasing with progressively
greater chemical `freedom'.  Since the statistical significance which
can be reached is satisfactory, it appears that the quark counting suffices
to describe the yields of heavier hadron resonances. 

\begin{table}[b]
\caption{ \label{fitsw}
Statistical parameters obtained from  fits of data for S--Au/W/Pb 
collisions at 200$A$ GeV,
without enforcing conservation of strangeness  \protect\cite{Let99}.
}
\begin{tabular}{ccccc}
\hline\hline
$\lambda_{ q}$&$\lambda_{ s}$&
$\gamma_{ s}$&$\gamma_{ q}$& $\chi^2/$dof\phantom{$\frac AB$} \\
\hline
                     1.52 $\pm$ 0.02
                 &   1$^*$
                 &   1$^*$
                 &   1$^*$
                 &   17  \\
                     1.52 $\pm$ 0.02
                 &   0.97 $\pm$ 0.02
                 &   1$^*$
                 &   1$^*$
                 &   18  \\
                     1.48 $\pm$ 0.02
                 &   1.01 $\pm$ 0.02
                 &   0.62 $\pm$ 0.02
                 &   1$^*$
                 &   2.4  \\
                    1.49 $\pm$ 0.02
                 &  1.00 $\pm$ 0.02
                 &  0.73 $\pm$ 0.02
                 &  1.22 $\pm$ 0.06
                 &  0.90\\  
\hline\hline
\hspace*{-3.5cm}$^*$ denotes fixed (input) values \phantom{$\frac{A}{A}$}\hspace*{-6cm}&&&&
\vspace*{-0.6cm}
\end{tabular}
\end{table}

We note, in the last raw in table \ref{fitsw}, that while the 
yield of strangeness is still suppressed as compared to expectations based
on absolute chemical equilibrium, the number of light quarks seen
exceeds the count expected. Another way to say the same thing
is that there is pion excess, or alternatively, entropy 
excess \cite{Let93,Let95,Gaz95}. In fact there is a maximum value that
$\gamma_q$ can assume: consider again Eq. (\ref{FDDis}) for pions,
assuming symmetry for $u,d$ quarks:
\begin{equation}\label{Bospi}
f_\pi\equiv {{d^6N_i^\pi}\over{d^3pd^3x}}= {3\over (2\pi)^3}
{1\over \gamma_q^{-2}e^{E_\pi/T}- 1}.
\end{equation}
We see that for 
\begin{equation}\label{gamlim}
\gamma_q\to e^{m_\pi/2T}=\gamma_q^c,
\end{equation}
the Bose condensation becomes possible. 

\begin{figure}[tb]
\vspace*{-1.1cm}
\centerline{\hskip 0.5cm
\epsfig{width=11.cm,clip=,angle=0,figure=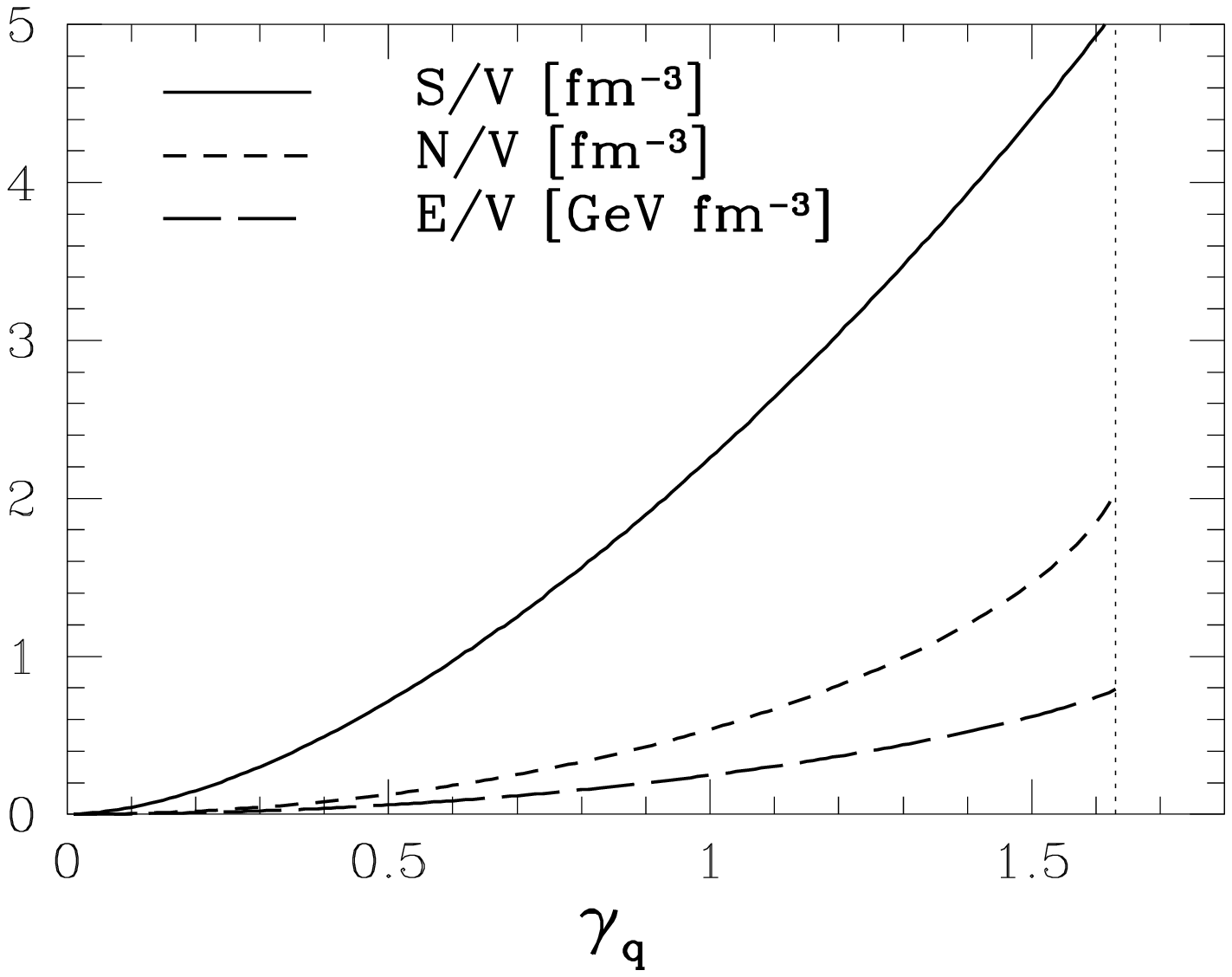}
}
\vspace*{-0.4cm}
\caption{ 
Pion-gas properties $N/V$ for particles, $E/V$ for energy, and 
$S/V$ for entropy  density, as functions of  $\gamma_{ q}$ 
at $T=142$\,MeV.
\label{abssne}
}
\end{figure}

The properties of 
the pion gas as function of $\gamma_q$ are shown in figure \ref{abssne},
where  the entropy, energy and particle number is expressed 
in terms of the momentum distribution function by:
\begin{eqnarray}\label{piprop} 
\frac S V&=&\int {d^3p}
       [(1+f_\pi)\ln(1+f_\pi)-f_\pi\ln f_\pi],\\
\frac E V&=&\int {d^3p}\sqrt{m_\pi^2+p^2}f_\pi,\\
\frac N V&=&\int {d^3p}f_\pi.
\end{eqnarray}
There is significant rise in entropy content as  $\gamma_q$ grows
toward the singular value. Excess entropy 
of a possibly color deconfined source can thus be squeezed in hadronization
into a hadron phase comprising an over saturated pion gas.
Indeed, when we perform an analysis of particles produced in 
the Pb--Pb 158\agev reactions, as well as at RHIC,  the value $\gamma_q^c$  
is always favored by a fit having a statistical 
significance \cite{Raf01}.

When an analysis of the Pb--Pb collision system is performed
allowing for quark pair 
abundance chemical non-equilibrium, we find that 
the light quark phase-space
occupancy parameter $\gamma_s$  
prefers the maximum allowed $\gamma_q^c$, and this is also the numerical
value preferred by the strange quark occupancy parameter $\gamma_s$. 
The freeze-out of particles is found at $T=150$ MeV. The velocity of 
(transverse) expansion $v_c$ (not further discussed here) is at half 
velocity of light. These results were obtained fitting all
particles shown in table \ref{resultpb2}.

\begin{table}[tb]
\caption{\label{resultpb2}
WA97 (top) and NA49 (bottom)  Pb--Pb 158\agev-collision hadron ratios
compared with phase-space fits.}
\begin{tabular}{lccll}
\hline\hline
\baselineskip 0.9cm
 Ratios & Reference &  Experimental data       & Pb$|^{ s,\gamma_{ q}}$ & Pb$|^{ \gamma_{q}}$\vphantom{$\frac AB$} \\
\hline
${\Xi}/{\Lambda}$ &  \protect\cite{Kra98} &0.099 $\pm$ 0.008                     & 0.096 & 0.095\\
${\overline{\Xi}}/{\bar\Lambda}$ &  \protect\cite{Kra98} &0.203 $\pm$ 0.024      & 0.197 & 0.199\\
${\bar\Lambda}/{\Lambda}$  &  \protect\cite{Kra98} &0.124 $\pm$ 0.013            & 0.123 & 0.122\\
${\overline{\Xi}}/{\Xi}$  &  \protect\cite{Kra98} &0.255 $\pm$ 0.025             & 0.251 & 0.255\\
\hline
${\rm K^+}/{\rm K^-}$         &  \protect\cite{Bor97}         &  1.800 $\pm$ 0.100         & 1.746  & 1.771 \\
${\rm K}^-/{\pi}^-$         &       \protect\cite{Sik99}&  0.082 $\pm$ 0.012        & 0.082  & 0.080\\
${\rm K}^0_{ s}/b$       &  \protect\cite{Jon96}   & 0.183 $\pm$ 0.027       & 0.192  & 0.195\\
${h^-}/b$                 &  \protect\cite{App99}     & \ 1.970 $\pm $ 0.100\ \      & 1.786  & 1.818 \\
$\phi/{\rm K}^-$   &  \protect\cite{Afa00}  & \ 0.145 $\pm$ 0.024\ \                 & 0.164  & 0.163 \\
${\bar\Lambda}/{ \bar p}$     &       $y=0$             &                       & 0.565  & 0.568 \\
\hline
 & $\chi^2$     &     & 1.6 & 1.15 \\
 &  $ N;\ p;\ r$     &    & 9;\ 4;\ 1& 9;\ 5;\ 1\\
\hline\hline
\vspace*{-0.6cm}
\end{tabular}
\end{table}

Due to their exceptional inverse slope, see figure \ref{PbPbtemp},
the yields of both $\Omega$ and $\overline\Omega$ are known,
to follow  different systematic behavior and have not been 
considered in this fit. With the parameters here determined
the  yields of $\overline\Omega$ are under predicted. This  excess 
yield originates at the lowest $m_\bot$, as we shall discuss  
below, see figure \ref{TdOK}.
The `failure' of a statistical-hadronization model to 
describe yields of soft $\Omega$ and $\overline\Omega$ has several 
possible explanations.  One is the possibility that an 
enhancement in production of $\Omega$ and $\overline\Omega$ is caused by 
pre-clustering of strangeness in the deconfined phase \bcite{Raf82a}.
This would enhance the production of all multistrange 
hadrons, but most prominently the highly phase-space-suppressed 
yields of $\Omega$ and $\overline\Omega$.  This mechanism 
would work only if pairing of strange quarks near to the phase 
transition were significant. Another possible mechanism of
$\Omega$ and $\overline\Omega$ enhancement is the distillation 
of strangeness \cite{Gre87,Raf87},
followed by breakup of strangeletts (strangeness enriched quark drops) which
could contribute to production of $\Omega$ and $\overline\Omega$.
The decay of disoriented chiral condensates has recently  been proposed as 
another source of soft  $\Omega$ and $\overline\Omega$ \bcite{Kap01}.

In view of these pre- and post-dictions of the 
anomalous yield of $\Omega$ and $\overline\Omega$, and the difference 
in shape of particle spectra, we believe that 
one should abstain from introducing these particles into 
statistical-hadronization-model fits.
We note that the early statistical descriptions of 
 yields of $\Omega$ and $\overline\Omega$ have not been sensitive to the
problems we described \bcite{Bec98,Let97b}. In fact, as
long as the parameter $\gamma_{ q}$ is not considered, it is not possible
to describe the experimental data at the level of precision that
would allow recognition of the excess yield of  $\Omega$ and $\overline\Omega$
within statistical hadronization model. For example, a  
chemical-equilibrium fit, which 
includes the yield of $\Omega$ and $\overline\Omega$,
has for 18 fitted data points with two parameters a 
$\chi^2/{\rm dof}=37.8/16$ \bcite{Bra99}. Such a fit is 
quite unlikely to contain all the physics, even if its 
appearance  to the naked eye  suggests that a very good 
description of experimental data as been achieved.

\begin{table}[tb]
\caption{\label{fitqpbs}
Upper section: the statistical model parameters
which best describe the experimental results for
Pb--Pb data seen in table\,\ref{resultpb2}.
Bottom section: energy per entropy, anti\-strangeness, and net strangeness
 of  the full hadron phase-space characterized by these
statistical parameters. In column two, we fix 
$\lambda_{ s}$ by the requirement of 
 conservation of strangeness.
}
\vspace{-0.2cm}
\begin{tabular}{lcc}
\hline\hline
 \vphantom{$\frac AB$}  & Pb$|_v^{ s,\gamma_{ q}}$ & Pb$|_v^{\gamma_{ q}}$\\
\hline
{\vphantom{$\frac AB$}}$T$ [MeV]             &  151 $\pm$ 3\phantom{00}      &  147.7 $\pm$ 5.6\phantom{00}           \\
$v_{\rm c}$           & 0.55 $\pm$ 0.05   & 0.52 $\pm$ 0.29        \\
$\lambda_{ q}$     & 1.617 $\pm$ 0.028 & 1.624 $\pm$ 0.029       \\
$\lambda_{ s}$     & 1.10$^*$         & 1.094 $\pm$ 0.02\phantom{0}        \\
$\gamma_{ q}$  & ${\gamma_{ q}^{\rm c}}^*=e^{m_{\pi}/(2T_{\rm f})}$ = 1.6  
&${\gamma_{ q}^{\rm c}}^*=e^{m_{\pi}/(2T_{\rm f})}$ = 1.6\\
$\gamma_{ s}/\gamma_{ q}$   & 1.00 $\pm$ 0.06  & 1.00 $\pm$ 0.06         \\
\hline 
$E/b$ [{GeV}]   &  4.0    &     4.1      \\
${s}/b$               & 0.70 $\pm$ 0.05  & 0.71 $\pm$ 0.05         \\
$E/S$ [{MeV}]   & 163 $\pm$ 1\phantom{00}    & 160 $\pm$ 1\phantom{00}           \\
$({\bar s}-s)/b\ \ $  & 0$^*$            &  0.04 $\pm$ 0.05 {\vphantom{$\frac AB$}}   \\ 
\hline\hline
$^*$ indicates values resulting from constraints.\phantom{$\frac{A}{A}$}\hspace*{-5.cm}&&
\end{tabular}
\end{table}

We also see, in table \ref{fitqpbs}, the yield of strangeness obtained in the 
fit of the final state hadron phase-space, which with 0.7 strange pairs
per baryon appears large. The question we next consider is what we
should have expected in hadronization of a deconfined quark matter phase. 
We consider the ratio of the equilibrium density 
of strangeness, arising in the
Boltzmann-gas limit,  to the baryon density 
in a fireball  of \QGP:
\beql{sdivb}
\frac{\rho_{ s}}{\rho_{ b}}={s\over b}={s\over q/3}=
\frac{\gamma_{ s}^{\mathrm{QGP}}}{\gamma_{ q}^{\mathrm{QGP}}} 
\frac{{(3/\pi^2)} T^3 W(m_{ s}/T)}
  {\frac23\!\left(\mu_{ q} T^2+{\mu_{ q}^3/ \pi^2}\right)}.
\eeq
 $W(x) =x^2K_2(x)$ defines in Boltzmann limit 
the equilibrium strange-quark density, 
with $g_{ s}=6$. We assume that to a first approximation,
perturbative thermal QCD corrections, cancel out in the ratio.
For $m_{ s}=200$ MeV and $T=150$ MeV, we have
\beql{sdivb1}
{s\over b}\simeq  
\frac{\gamma_{ s}^{\mathrm{QGP}}}{\gamma_{ q}^{\mathrm{QGP}}} 
   \frac{0.7}{\ln \lambda_{ q} +{{(\ln \lambda_{ q})^3}/{\pi^2}}}.
\eeq

\begin{figure}[tb]
\centerline{\hskip 0.5cm
\epsfig{width=11.cm,clip=,angle=0,figure=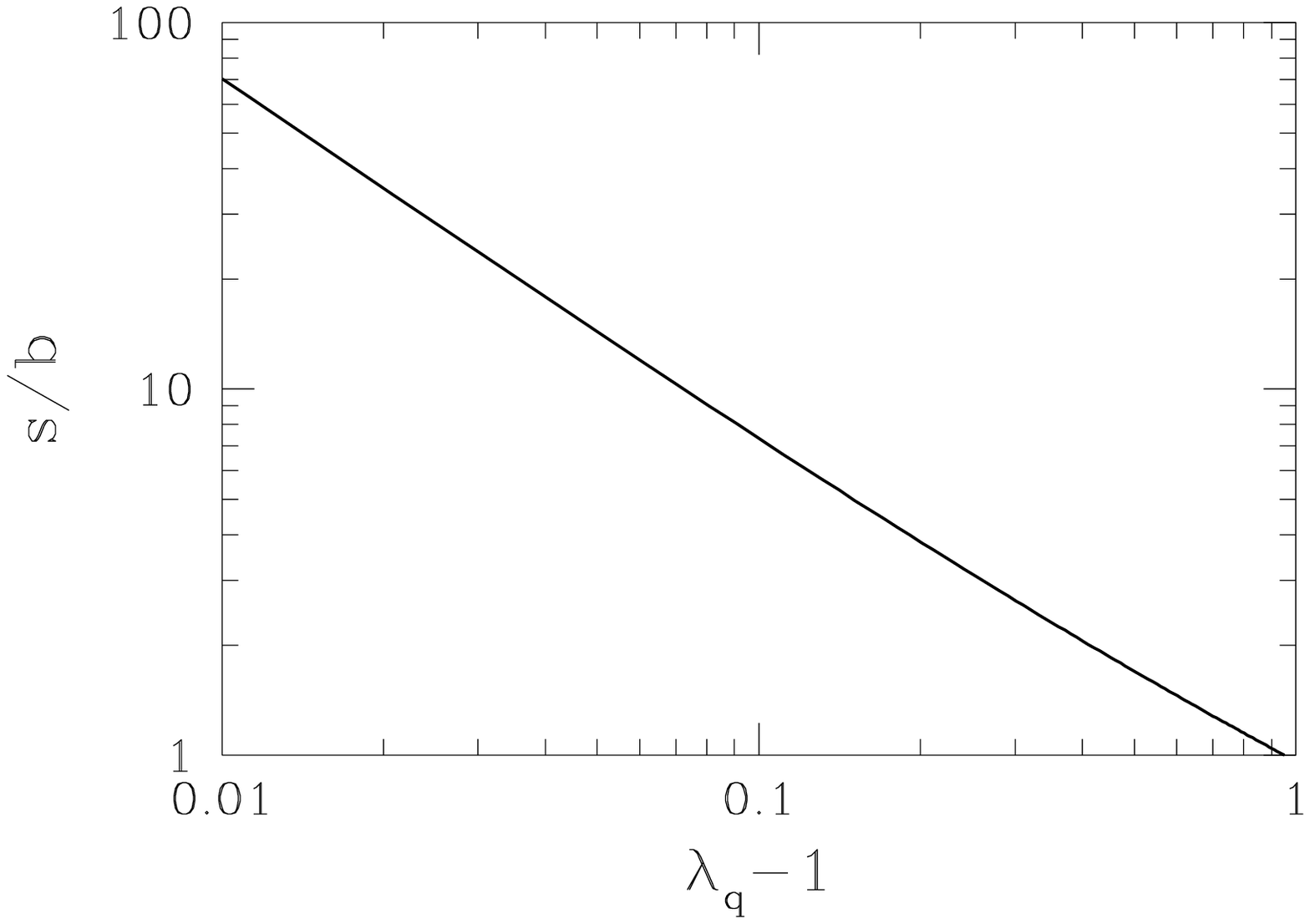}
}
\caption{ 
The yield of strangeness per baryon as a function of $\lambda_{ q}$ 
in equilibrated \QGP.
\label{PLSBLAMQ}
}
\end{figure}
The relative yield $s/b$ is mainly dependent on the value of $\lambda_{ q}$. 
In the approximation considered, it is nearly temperature-independent,
the result is shown in figure \ref{PLSBLAMQ}, as a function of 
$\lambda_{ q}-1$ (the variable chosen to enlarge the interesting region
$\lambda_{ q}\to 1$) for 
$ \gamma_{ s}^{\mathrm{QGP}}=\gamma_{ q}^{\mathrm{QGP}}=1$. 
At  the top SPS energy where $\lambda_{ q}\to 1.5$--1.6, we see that 
the equilibrium yield is at 1.5 pairs of strange quarks 
per participating baryon. Considering the experimental yield 
in table \ref{fitqpbs}, we thus conclude 
$\gamma_{ s}^{\mathrm{QGP}}\sim 0.5$.
Experimentally  the directly accessible observable is the occupancy of the 
 hadron-strangeness phase-space and we show in table \ref{fitqpbs}, 
$\gamma_{ s}^{\mathrm{HG}}\simeq 1.6 $. This thus implies that
in the hadronization of a presumed QGP the smaller phase-space of
hadrons is overpopulated by a QGP abundance, even when it is 
originally in the QGP phase well below 
the equilibrium value. 

The above analysis demonstrated  the importance  of
the study of quark-pair abundance chemical 
non-equilibrium features and how these
are capable to refine our understanding of the 
QGP formation and hadronization.

\section{QGP and chemical analysis}

An interesting feature seen in table \ref{fitsw} is that the
value of $\lambda_s$ converges to unity. This is not at all what 
could be expected for a equilibrated hadron gas. Namely, the 
value of $\lambda_s$ arises from the requirement of 
conservation of strangeness. Specifically the net strangeness is given by,
 \beqar
0=  \langle n_{ s}\rangle  - \langle n_{ \bar s}\rangle= { {T^3} \over {2\pi^2}} \Bigl[ 
   &&\hspace{-0.6cm}  (\lambda_{ s} \lambda_{ q}^{-1} -
    \lambda_{ s}^{-1} \lambda_{ q}) \gamma_{ s}\gamma_{ q} F_{\mathrm K}
  +\,
     (\lambda_{ s} \lambda_{ q}^{2} -
    \lambda_{ s}^{-1} \lambda_{ q}^{-2}) \gamma_{ s} \gamma_{ q}^2 F_{\mathrm Y}
\nonumber  \\
&&\hspace{-0.6cm}
  +\,
   2(\lambda_{ s}^2 \lambda_{ q} - 
     \lambda_{ s}^{-2}\lambda_{ q}^{-1}) \gamma_{ s}^2\gamma_{ q} F_\Xi 
  +\,
    3(\lambda_{ s}^{3} - 
  \lambda_{ s}^{-3})\gamma_{ s}^3 F_\Omega \,     \Bigr] ,
 \eeqarl{KMR9explicit}
where we have employed the phase-space integrals for known hadrons,
 \beqarl{4b}
   F_{\mathrm K} &&\hspace{-0.6cm}=  \sum_j g_{{\mathrm K}_j} W(m_{{\mathrm K}_j}/T);\
           {\mathrm K}_j={\mathrm K},{\mathrm K}^\ast,{\mathrm K}_2^\ast,\ldots \!,\ \
           m\le 1780 \ {\mathrm{MeV}}\, ,
   \nonumber\\
   F_{\mathrm Y} &&\hspace{-0.6cm}=  \sum_j g_{\mathrm Y_{\!j}} W(m_{\mathrm Y_{\!j}}/T);\
           \mathrm Y_{\!j}=\Lambda, \Sigma, \Sigma(1385),\ldots \!,\ 
           m\le 1940\ {\mathrm{MeV}}\, ,
   \nonumber\\
   F_\Xi &&\hspace{-0.6cm}=  \sum_j g_{\Xi_j} W(m_{\Xi_j}/T);\
           \Xi_j=\Xi,\Xi(1530),\ldots \!,\ \
           m\le 1950\ {\mathrm{MeV}}\, ,
   \nonumber\\
   F_\Omega &&\hspace{-0.6cm}=  \sum_j g_{\Omega_j}
W(m_{\Omega_j}/T);\            \Omega_j=\Omega,\Omega(2250)\, .
 \eeqar
The $g_i$ are the spin--isospin degeneracy factors, $W(x)=x^2K_2(x)$,
where $K_2$ is the modified Bessel function.

In general, \req{KMR9explicit} must be equal to zero since strangeness
is a conserved quantum number with respect to the strong
interactions, and no strangeness is brought into the reaction.
The possible exception is dynamic evolution with asymmetric 
emission of strange and antistrange hadrons \cite{Raf87}. 
\req{KMR9explicit} can be solved analytically when the contribution of 
multistrange particles is small:
\beql{KMR10}
{\lambda_{ s}|_0} =\lambda_{ q} \sqrt
{
{{ F_{\mathrm K} +  \gamma_{ q}\lambda^{-3}_{ q} F_{\rm Y}} \over 
{ F_{\mathrm K} +  \gamma_{ q}\lambda^3_{ q} F_{\rm Y}}} 
}    .
\eeq 
We thus see that except for a very exceptional point where the kaon and
hyperon strangeness phase-spaces for a given value of baryochemical
properties are of same magnitude, the value of $\lambda_s$ will
not be unity. One can of course force the hadron multiplicity
description to this value, but the prize one pays is a greatly reduced
statistical significance \cite{Bra99,Bra96d}, where chemical
equilibrium has been assumed. We note that for the value $\lambda_s=1$, we
can analytically solve \req{KMR9explicit}, including the effect
of multistrange hadrons and obtain:
\beql{s0lams1}
\mu_{ b}
=3T\,\ln(x+\sqrt{x^2-1}),\qquad 
1\le x=\frac{F_{\mathrm K}-2\gamma_{ s}F_\Xi}{2\gamma_{ q}F_{\mathrm Y}}.
\eeq

In what situation is the value $\lambda_s\to 1$ natural?
This value will arise when for all baryo-chemical conditions 
the strange and antistrange quark numbers can balance independently.
This will in most cases be in the phase in which strange quarks can
roam freely. In this case we have instead of condition \req{KMR9explicit},
\beqarl{difss}
0&\hspace{-0.3cm}=&\hspace{-0.3cm}\langle n_{ s}\rangle-\langle n_{\bar{ s}}\rangle \nonumber\\&\hspace{-0.3cm}=&\hspace{-0.3cm}
g\!\int\!\!\frac{d^3p}{(2\pi)^3}\!
\!\left(\!\frac1{\!1+\gamma^{-1}_{ s}\lambda_{ s}^{-1}
\exp\!\left({{\sqrt{p^2+m^2_{ s}}}\over {T}}\right)}
    -\ \frac1{\!1+\gamma^{-1}_{ s}\lambda_{ s} 
\exp\!\left({{\sqrt{p^2+m^2_{ s}}}\over {T}}\right)}\!\right)\!.
\eeqar
We note the change in the power of $\lambda_{ s}$ between these two terms, and recognize
that this integral can vanish only for $\lambda_{ s}\to 1$. 

There is potentially small but
 significant asymmetry in $\lambda_{s}$ due to the Coulomb 
charge present in baryon-rich quark matter: long-range electromagnetic
potential $V_{\rm C}\ne 0$ influence strange and antistrange particles differently, 
and  a slight deviation $\lambda_{ s}>1$ is needed in order to 
compensate for this effect in the QGP phase. 
We have 
as generalization of \req{difss} \bcite{Let99c},
\beqarl{Nsls}\nonumber
0&\hspace{-0.3cm}=&\hspace{-0.3cm}\langle N_{ s} \rangle-\langle N_{ \bar s}\rangle  \nonumber\\&\hspace{-0.3cm}=&\hspace{-0.3cm}
g\! \int\limits_{R_{\rm f}}\! \frac{d^3r\,d^3p}{V(2\pi)^3}
\left(\!\!
 \frac1{1+\gamma_{ s}^{-1}\lambda_{ s}^{-1}e^{(E(p)-\frac13 V_{\rm C}(r))/T}}
 -\frac1{1+\gamma_{ s}^{-1}\lambda_{ s}e^{(E(p)+\frac13 V_{\rm C}(r))/T}}
 \!\!\right)\!,
\eeqar
which clearly cannot vanish for $V_{\rm C}\ne 0$, in the limit $\lambda_{ s}\to1$.
The volume integral is here over the fireball of size $R_{\rm f}$. In the
Boltzmann approximation one easily finds that 
\beql{tilams}
\tilde\lambda_{ s}\equiv \lambda_{ s} \ell_{\rm C}^{1/3}=1,\qquad
\ell_{\rm C}\equiv
\frac{\int_{R_{\rm f}} d^3r \,e^{V/{T}}} {\int_{R_{\rm f}} d^3r}.
\eeql{lamQ}
 $\ell_{\rm C}<1$  expresses the Coulomb deformation of 
strange quark phase-space. $\ell_{\rm C}$ is not a fugacity that 
can be adjusted to satisfy a chemical condition,
since consideration of $\lambda_i,\ i= u,\,d,\,s$, exhausts all available
chemical balance conditions for the abundances of hadronic particles,
and allows introduction of the fugacity associated with the Coulomb
charge of quarks and hadrons. 
Instead,  $\ell_{\rm C}$ characterizes the distortion of the phase-space 
by the long-range Coulomb interaction.
This Coulomb distortion of the quark phase-space is 
naturally also present for $u$, $d$ quarks, but appears 
less significant given that  $\lambda_{ u}$ and 
 $\lambda_{ q}$ are empirically determined. On the 
other hand, this effect eliminates much if not all the difference between
$\mu_u$ and $\mu_d$ we have described above, since the quark abundance asymmetry
arises naturally due to Coulomb effect --- said differently, the Coulomb
effect deforms the phase-space such that it is natural to have more 
$d$ than $u$ quarks and thus the  asymmetry between  $\lambda_d$ and $\lambda_u$ 
is reduced. 

Choosing  $T=140$ MeV and 
$m_{ s}=200$ MeV, and noting that the value of $\gamma_{ s}$ 
is practically irrelevant
since this factor cancels out in the Boltzmann approximation
we find for $Z_{\rm f}=150$ that the value 
$\lambda_{ s}=1.10$ is needed for $R_{\rm f}=7.9$\,fm, 
whereas for S--Au/W/Pb reactions,  similar analysis leads to a value 
$\lambda_{ s}=1.01$. Chemical
freeze-out at higher temperature, \eg, $T=170$ MeV, leads for 
$\lambda_{ s}=1.10$ to somewhat 
smaller radii, which is consistent with the higher temperature used.

The fit of the Pb--Pb system seen in table \ref{fitqpbs} 
converged just to the value expected if QGP were formed. However, 
accidentally for $\gamma_q=\gamma_q^{\rm c}$ this happens
also to be where in hadronic gas strangeness
balance is found. Thus in case of Pb--Pb the indication from
chemistry that a QGP has been formed arises mainly from the fact that
a significant entropy excess is available,
(light quark-pair excess) and the strange quark-pair excess, seen 
the relatively high value of
$\gamma_s\simeq \gamma_q$ in table \ref{fitqpbs}.
When compared to the yield expected from QGP, this  implies that 
$\gamma_{ s}^{\mathrm{HG}}/ \gamma_{ s}^{\mathrm{QGP}}\simeq 3$.
To better understand this, we compare the
phase-space of strangeness in \QGP with that of the resulting hadronic gas. 
The absolute strangeness yields must be the same in both phases. We perform
the comparison assuming that in a fast hadronization of QGP neither temperature
$T$, nor the baryochemical potential $\mu_b$, nor the reaction volume 
have time to  change, the difference in the properties of the phases
is absorbed in the change in the occupancy of phase-space, $\gamma_i$
in the two phases. 

We relate the two phase-space occupancies in 
HG and QGP, by equating the strangeness content in the two  phases.
One has to keep in mind that there is some additional production
of strangeness due to gluon hadronization, however, this is
not altering the argument presented significantly. 
On canceling out the common normalization factor $T^3/(2\pi^2)$, 
we obtain 
\beql{stranHAD}
\gamma_{ s}^{\mathrm{QGP}} V^{\mathrm{QGP}}g_{ s} 
W\!\left({m_{ s}\over T^{\mathrm{QGP}}}\right)\simeq
\gamma_{ s}^{\mathrm{HG}}  V^{\mathrm{HG}} \!
\left({ \gamma_{ q}\lambda_{ q}\over \lambda_{ s}}F_{\rm K}
          \!+\!{\gamma_{ q}^2\over \lambda_{ q}^{2}\lambda_{ s}}F_{\rm Y}\right)\!.
\eeq

Here we have, without loss of generality, 
  followed the $\bar s$-carrying hadrons in the hadronic gas
 phase-space, and we have omitted the contribution of multistrange
antibaryons for simplicity. We now use the condition that strangeness is
conserved, \req{KMR10}, to eliminate $\lambda_{ s}$ from \req{stranHAD}, 
and obtain (making explicit which statistical parameter occurs in which phase)
\beql{stranHAD1}
{\gamma_{ s}^{\mathrm{HG}}\over \gamma_{ s}^{\mathrm{QGP}}} 
{V^{\mathrm{HG}}\over V^{\mathrm{QGP}}}
  =
{{g_{ s} W(m_{ s}/T^{\mathrm{QGP}})}\over{ \sqrt{
(\gamma_{ q} F_{\rm K}+ \gamma_{ q}^2 \lambda_{ q}^{-3}F_{\rm Y})
(\gamma_{ q} F_{\rm K}+ \gamma_{ q}^2 \lambda_{ q}^{3}F_{\rm Y})}}}.
\eeq

\begin{figure}[tb]
\vspace*{-1.1cm}
\centerline{\hskip 0.5cm
\epsfig{width=11.cm,clip=,angle=0,figure=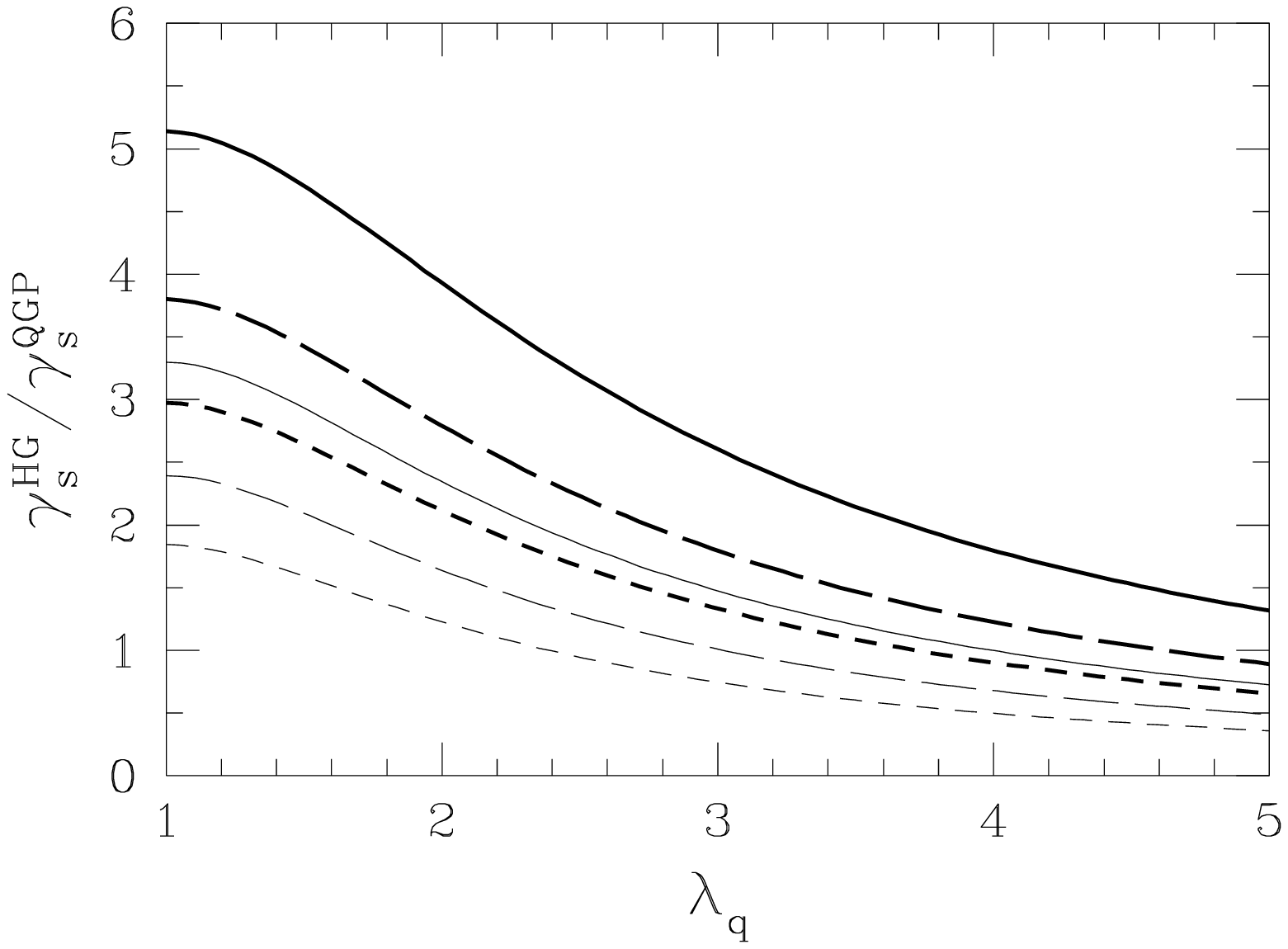}
}
\vspace*{-0.4cm}
\caption{ 
The HG/QGP  strangeness-occupancy $\gamma_{ s}$ ratio in sudden hadronization
as a function of  $\lambda_{ q}$. Solid lines, $\gamma_{ q}^{\rm HG}=1$; 
long-dashed lines, $\gamma_{ q}^{\rm HG}=1.3$; and short-dashed lines, 
$\gamma_{ q}^{\rm HG}=1.6$.
Thin lines are for $T=170$ and thick lines for $T=150$ MeV, for both 
phases. 
\label{PLGAHGQGP}
}
\end{figure}

This ratio is shown in figure \ref{PLGAHGQGP}, and the two lines of particular
interest are the thin solid line (chemical equilibrium
$\gamma_{ q}^{\rm HG}=1$ with $T=170$ MeV in 
hadronic gas phase), and the short dashed thick line 
($\gamma_{ q}^{\rm HG}=1.6$ at $T=150$ MeV) which correspond to 
the two hadronization scenarios which can be used to fit the SPS 
experimental results. We see that, near to the established value 
$\lambda_q\simeq 1.6$, both yield a squeeze by factor 3 for the ratio
of the strangeness phase-space occupancies. 

This result implies that hadronization of QGP is 
at SPS and RHIC accompanied by an increase by factor
three in the value of $\gamma_s$ and thus it is most 
important to allow in the study of particle yields for
this of-equilibrium property, otherwise the tacit
assumption has been made that QGP has not been formed. 
 If at RHIC 
conditions are reached in which strangeness in QGP
is closer to absolute chemical equilibrium, 
$\gamma_s^{\rm QGP}\to 1$, even a more
significant overpopulation by strangeness of the
hadronic gas phase would result, $\gamma_s^{\rm HG}\to 3$.

\section{Thermal freeze-out and single freeze-out}\label{thermal}
The non-equilibrium features introduced into the
chemical analysis  reduce the 
chemical freeze-out temperature to values close
to those expected for the thermal freeze-out, that is when
the spectral shape of particles freezes out after at
a low particle density also
elastic hadron interactions cannot occur. One is thus tempted
to ask the question if both freeze-out conditions
(chemical and thermal) are not actually one and the same. 
The confirmation came when we found 
 that the spectra and yields of hyperons, antihyperons
 and kaons were very well described within this scenario. However, 
we could not publish the results of a  single freeze-out scenario 
at SPS~\cite{Raf99a}, our scientific findings disagreed  with
personal  opinion of  than associate editor of The Physical  Review Letters,
Herr Ulrich  Heinz. Our submission to PRL (dated March 8, 1999)
occurred on the same day as the submission to Physics Letters B 
of the two-freeze-out chemical equilibrium, however this work 
was published~\cite{Bra99}, which as history has often shown makes a lot
of difference.

Despite this setback we proceeded to enlarge on our single-freeze-out
sudden hadronization picture for SPS, which in our opinion was 
strongly supported by many experimental facts,
and we could present these results at  meetings~\cite{Raf99}. 
Specifically, the spectra of strange hadrons and anti-hadrons show universal slope,
see figure  \ref{PbPbtemp}, which supports such a single freeze-out 
scenario decisively. The understanding  of pion spectra is very difficult since 
many (yet undiscovered) resonances contribute in a relevant fashion, 
and the soft part 
of the spectrum is particularly prone to `deformation' by
hadronization mechanisms and unknown matter flows~\cite{Let01}. 
Thus we report  here on our precise study of spectra of 
kaons and hyperons~\cite{Tor01}.

The final particle distribution is composed of directly 
produced particles and decay products of heavier hadronic resonances:
\begin{equation}\label{2body}
\frac{dN_X}{dm_\bot} =\left.  \frac{dN_X}{dm_\bot}\right|_{\rm direct} +
\sum_{ \forall R \rightarrow X + 2+\cdots } 
\left.\frac{dN_X}{dm_\bot}\right|_{R \rightarrow X + 2 +\cdots}.  
\end{equation}
$R(M,M_T,Y) \rightarrow X(m,m_T,y)+2(m_2)+\cdots$, where we 
indicate by the arguments that only for the decay  particle  $X$ 
we keep the information about the shape of the
momentum spectrum. 

In detail, the decay contribution to yield of 
$X$ is:
\begin{equation}
\label{reso}
\frac{dN_X}{d {m^2_\bot} d y }=
\frac{g_{r} b}{4 \pi p^{*}}
\int_{Y_-}^{Y_+}\!\! dY
\int_{M_{T_-}}^{M_{T_+}} dM_{T}^{2}\, 
\frac{M}{\sqrt{P_{T}^2 p_{T}^2 -\{M E^{*} - M_{T} m_{T} \cosh\Delta Y\}^2}}
\,\frac{d^2 N_{R}}{dM_{T}^{2} dY},  
\end{equation}
We have used $\Delta Y=Y-y$, and
$\sqrt{s}$ is the combined invariant mass of the 
decay products other than particle $X$
and $E^{*}=(M^2-m^2-m_2^2)/2M$, $p^{*}=\sqrt{E^{*2}-m^2}$ 
are the energy, and momentum, of the decay
particle $X$ in the rest frame of its parent.
The limits on the integration are the maximum values accessible
to the decay product $X$:
\[
Y_{\pm}=\hspace*{-0.2cm}y \pm \sinh^{-1}\!\left(\frac{p^{*}}{m_{T}}\right), 
\qquad
M_{T_{\pm}}=M \frac{E^{*} m_{T} \cosh\Delta Y \pm p_{T} 
\sqrt{p^{*2}-m_{T}^{2} \sinh^{2} \Delta Y}}
{m_{T}^{2} \sinh^{2} \Delta Y+m^{2}}.
\]

The theoretical primary particle
spectra (both those directly produced and parents of 
decay products) are derived from the Boltzmann 
distribution by Lorenz-transforming from a flowing 
intrinsic fluid element to the CM-frame, and 
integrating over allowed 
angles between particle direction  and local flow.

We introduce in the current analysis 
two velocities: a local  flow velocity $v$ of the fireball 
matter where from particles emerge,
and hadronization surface (breakup) velocity which we refer to 
as $v_{\rm f}^{\,-1}\equiv dt_{\rm f}/dx_{\rm f}$.  Particle production is controlled by the 
effective volume element, which comprises this quantity. In detail: 
\begin{equation}\label{v2v}
d S_{\mu} p^{\mu} = 
  d \omega \left(1- \frac{\vec v_{\rm f}^{\,-1} \cdot \vec p}{E}\right),
\qquad  d\omega \equiv \frac{d^3xd^3p}{(2\pi)^3}.
\end{equation}
The  Boltzmann distribution we adapt has thus the form
\begin{equation}
\frac{d^2 N}{dm_{T} dy} \propto
\left(1- \frac{\vec v_{\rm f}^{\,-1} \cdot \vec p}{E}\right)
\gamma\,  m_{T} \cosh y \,
e^{-\gamma \frac E T \left(1-\frac{\vec v\cdot \vec p}{E}\right)},
\end{equation}
where $\gamma=1/\sqrt{1-v^2}$\,.
The normalization for each hadron type $h=X,R$ is
$N^h = V_{\rm QGP} \prod^{n}_{i\in h} \lambda_{i} \gamma_{i}$. 
We use the chemical parameters 
$\lambda_{i}$ and $ \gamma_{i}$, $i=q,s$,
as obtained in the chemical analysis.

The experimental data we consider are  published 
$m_\bot$ distribution \cite{Ant00},  
with additional information obtained about absolute normalization.
This allowed us to perform the spectral  shape analysis
together with yield analysis, which indicates that 
the reaction volume is increasing as expected 
with the centrality of the 
reaction. This said, we will only focus here on 
the question if the shape of the spectra is consistent with 
the chemical freeze-out condition~\cite{Tor01}. 
The best fit to the spectra in fact produces the temperature and
transverse velocity in excellent agreement with those inferred from
chemical analysis we have discussed above. 

\begin{figure}[tb]
\vspace*{-1.1cm}
\centerline{\hskip 0.1cm
\epsfig{width=7.5cm,clip=,angle=0,figure=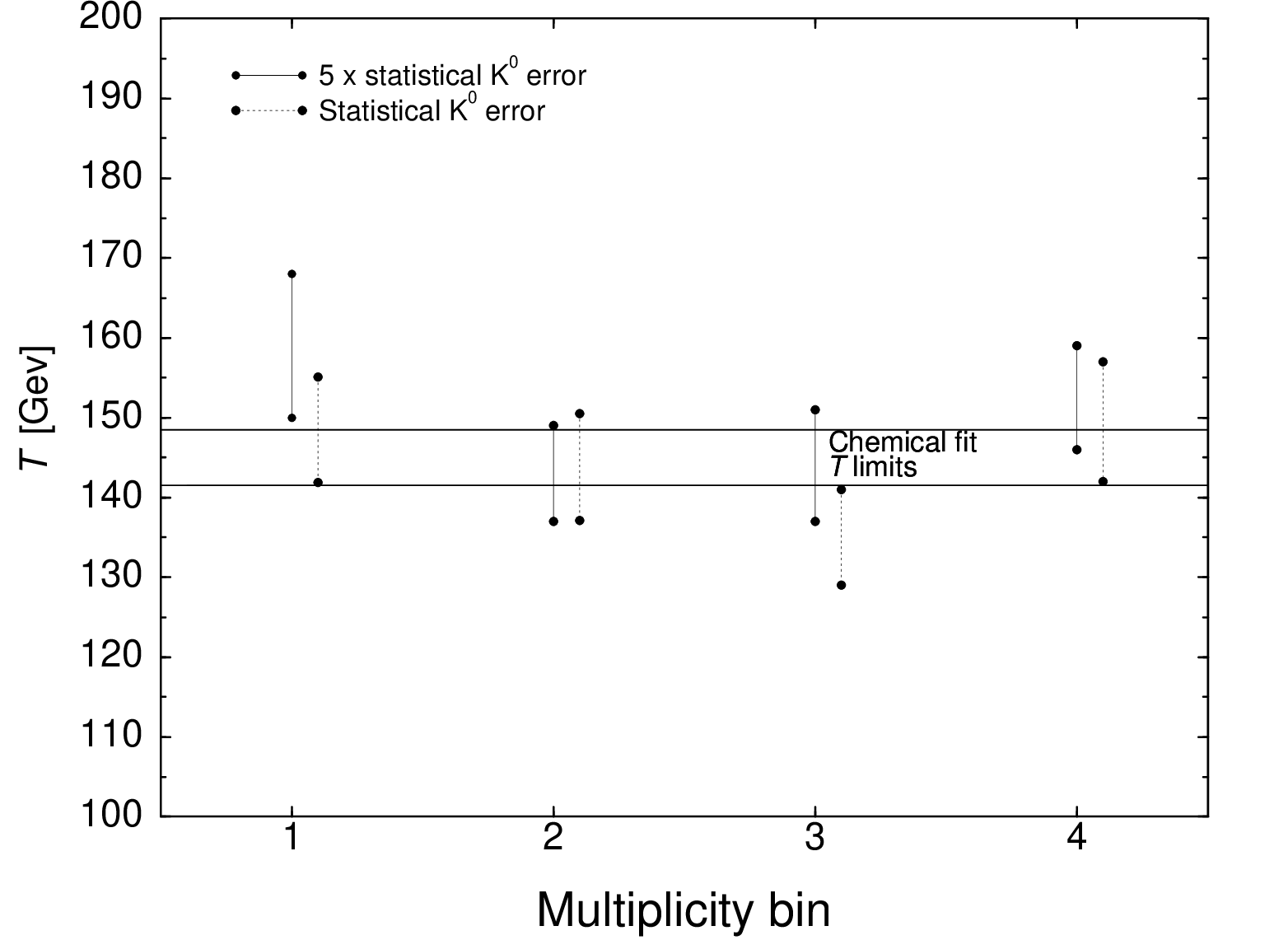}
\hspace*{.1cm}
\epsfig{width=7.1cm,clip=,angle=0,figure=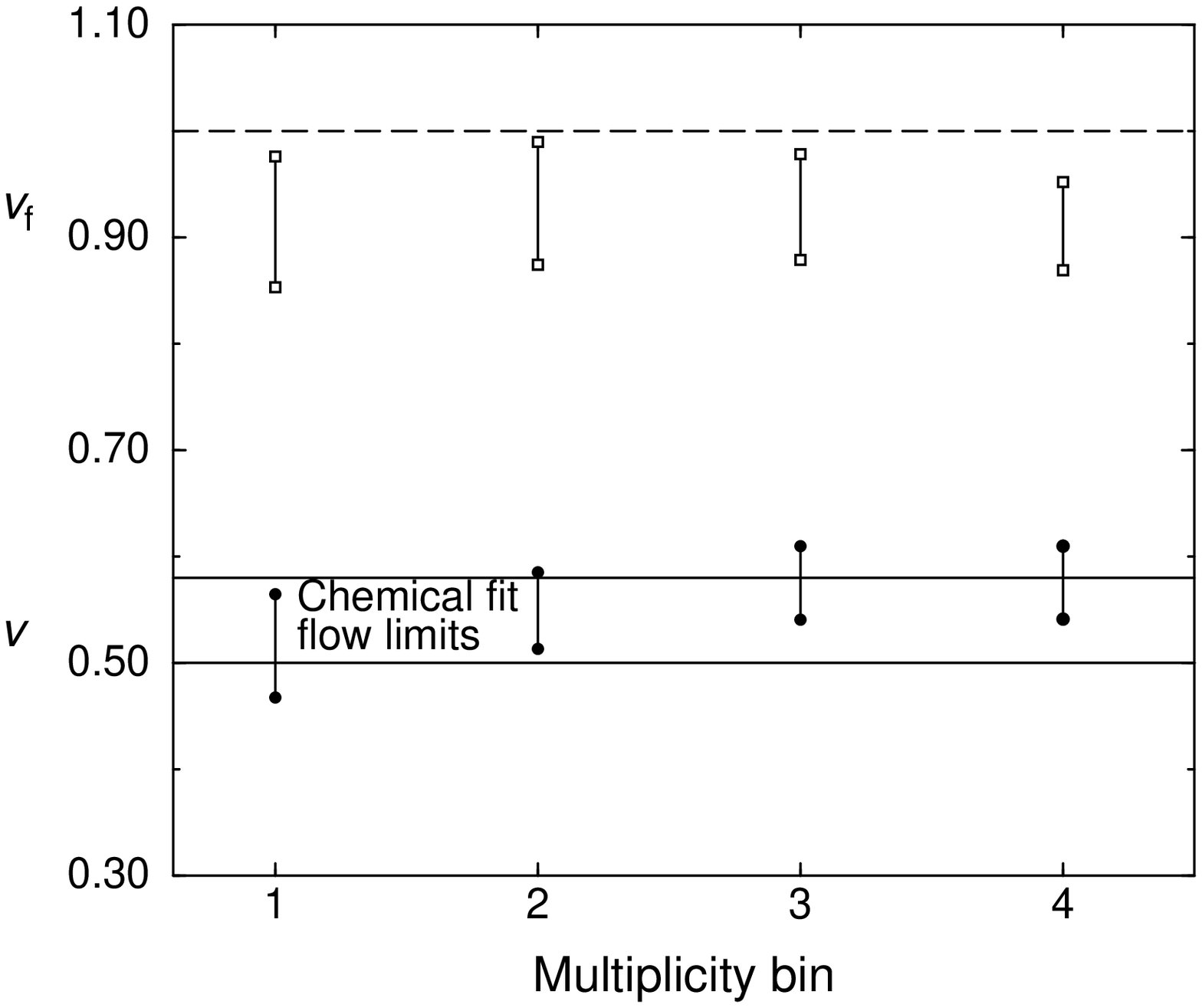}
}
\vspace*{-0.4cm}
\caption{ 
The thermal freeze-out temperature $T$ (left), 
flow velocity $v$ (bottom right), and
breakup (hadronization hyper-surface-propagation) velocity $v_{\rm f}$ 
(top right) for various collision-centrality bins. 
The upper limit $v_{\rm f}=1$ (dashed line) and 
chemical-freeze-out-analysis limits for $v$ (solid lines) are also shown.
For the temperature, results obtained with increased error for kaon 
spectra are also shown.
\label{TdTTdv1v2} 
}
\end{figure}

We  show,  in figure \ref{TdTTdv1v2}, the parameters
determining the shape of the $m_\bot$ distributions,
that is $T,v,v_{\rm f}$,
as function of the centrality for scattering 
bin 1, 2, 3, 4 with the most
central bin being 4.  The horizontal lines 
delineate range of result of the chemical analysis.

There is 
no indication of a significant, or systematic, change of $T$ with centrality.
This is consistent with the believe that the physics we 
are considering arises  in all centrality bins explored by the 
experiment WA97 in Pb--Pb reactions at 158\agev, 
\ie, for the number of participants greater than 60. 
Only most peripheral interactions  
produce a change in the pattern of strange hadron production~\cite{Kab99,Kab99b},
and we are anticipating with deep interest the more peripheral hyperon
spectra which should become available from experiment NA57.  
The magnitudes of the 
collective expansion velocity $v$  and the  break-up (hadronization) speed 
parameter $v_{\rm f}$ also do not show dependence on centrality,
though within the experimental error, one could argue 
inspecting  figure \ref{TdTTdv1v2} (right) that there is 
systematic increase in transverse flow velocity $v$ with centrality and thus 
size of the system. Such an increase
is expected, since the more central events comprise 
greater volume of matter, which allows more time for development
of the flow.  Interestingly, it is in $v$, and not in $T$, that the 
slight change of spectral slopes noted in the presentation of the 
experimental data~\cite{Ant00} is found.

The value of the beak-up (hadronization) speed 
parameter $v_{\rm f}$ shown in the top portion (right) of 
figure \ref{TdTTdv1v2} is near to 
velocity of light which is consistent with the picture of a 
sudden breakup of the  fireball. This 
hadronization surface velocity $v_{\rm f}$ was in the chemical
fit fixed to be equal to $v$, as there was not enough sensitivity in
purely chemical fit to  determine the value of $v_{\rm f}$.

\begin{figure}[p]
\begin{tabular}{lr}
\vspace*{1cm}
\epsfig{width=5.8cm,clip=,angle=-90,figure=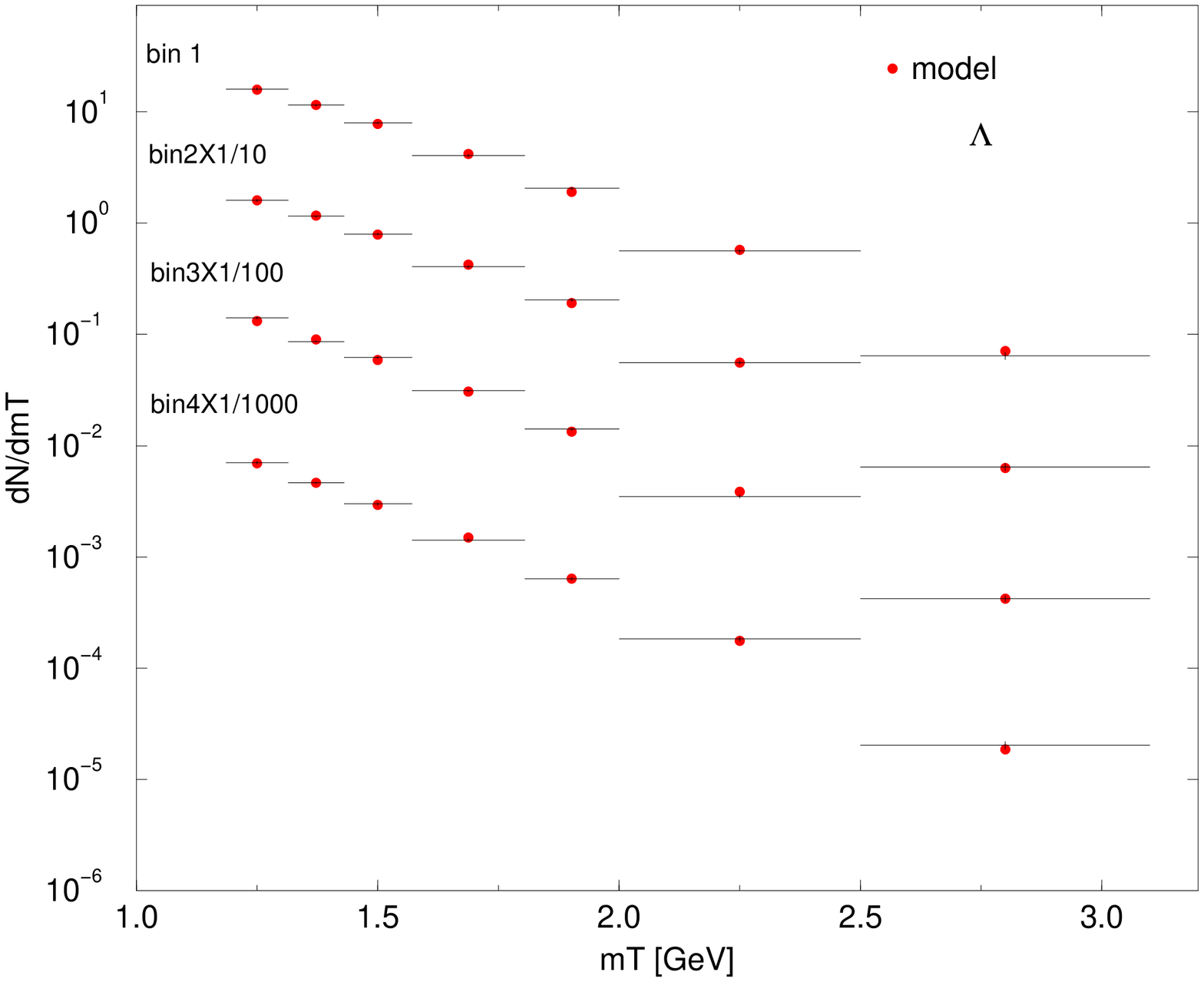}
&\epsfig{width=5.8cm,clip=,angle=-90,figure=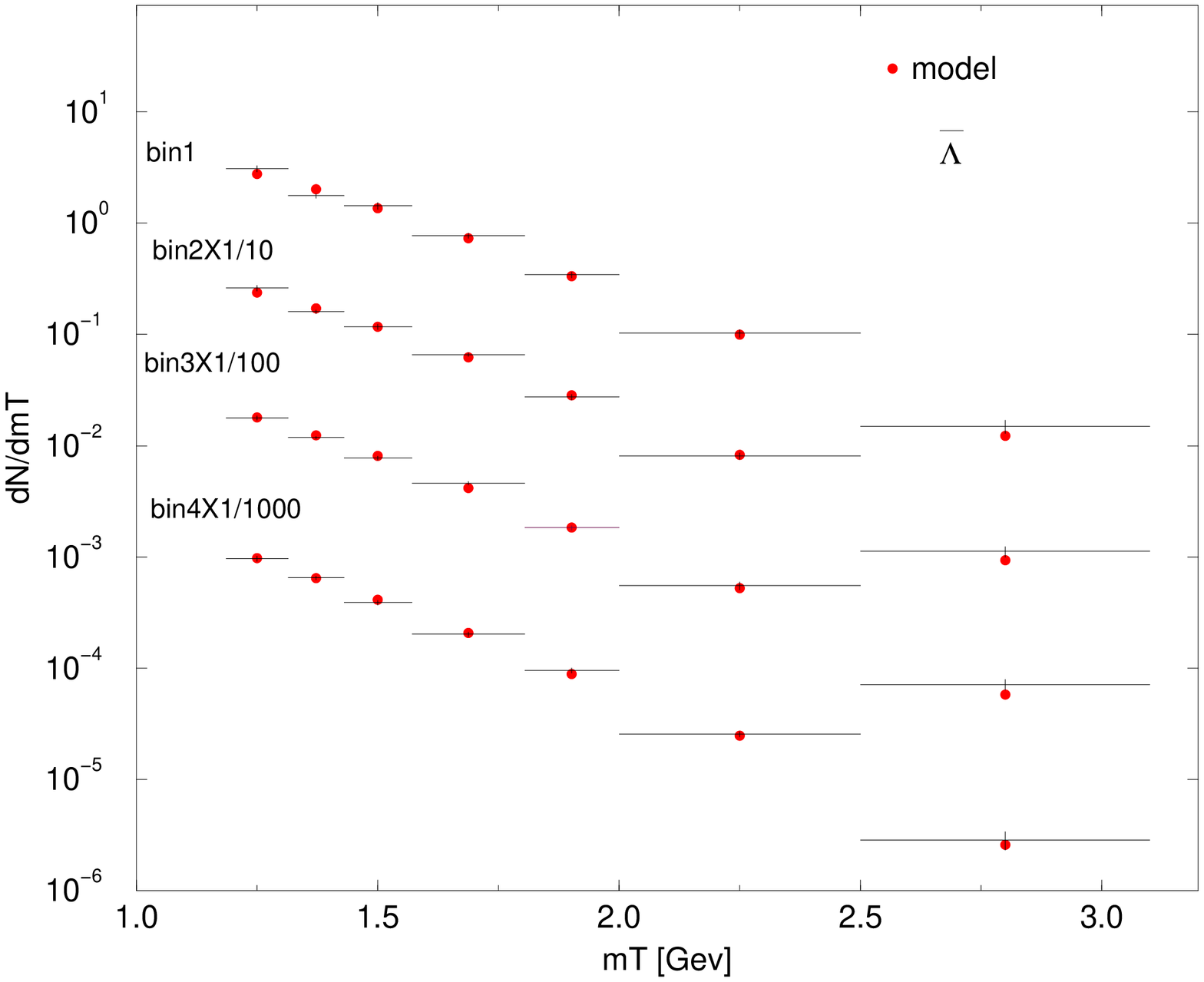}\\
\epsfig{width=5.8cm,clip=,angle=-90,figure=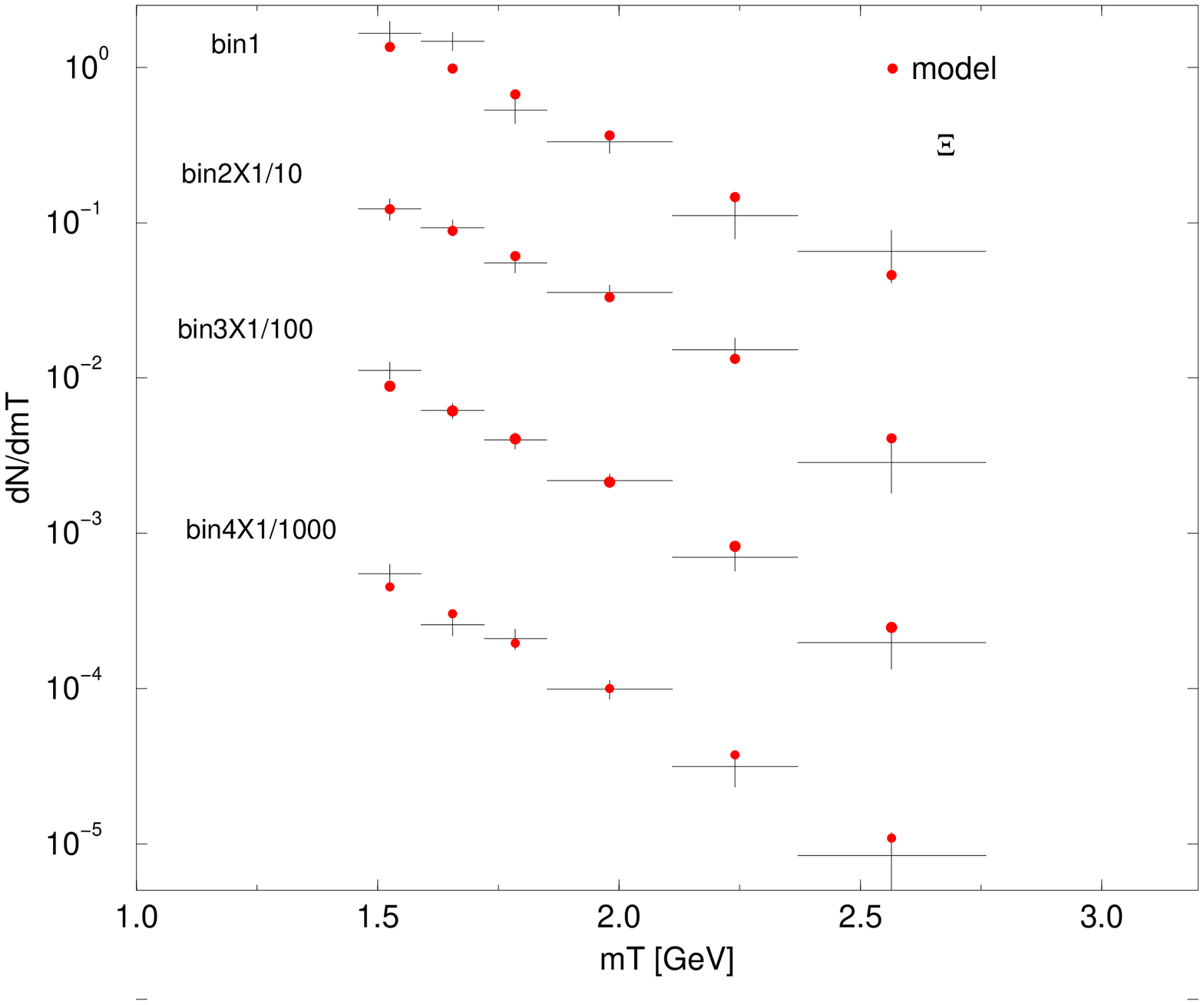}
&\epsfig{width=5.8cm,clip=,angle=-90,figure=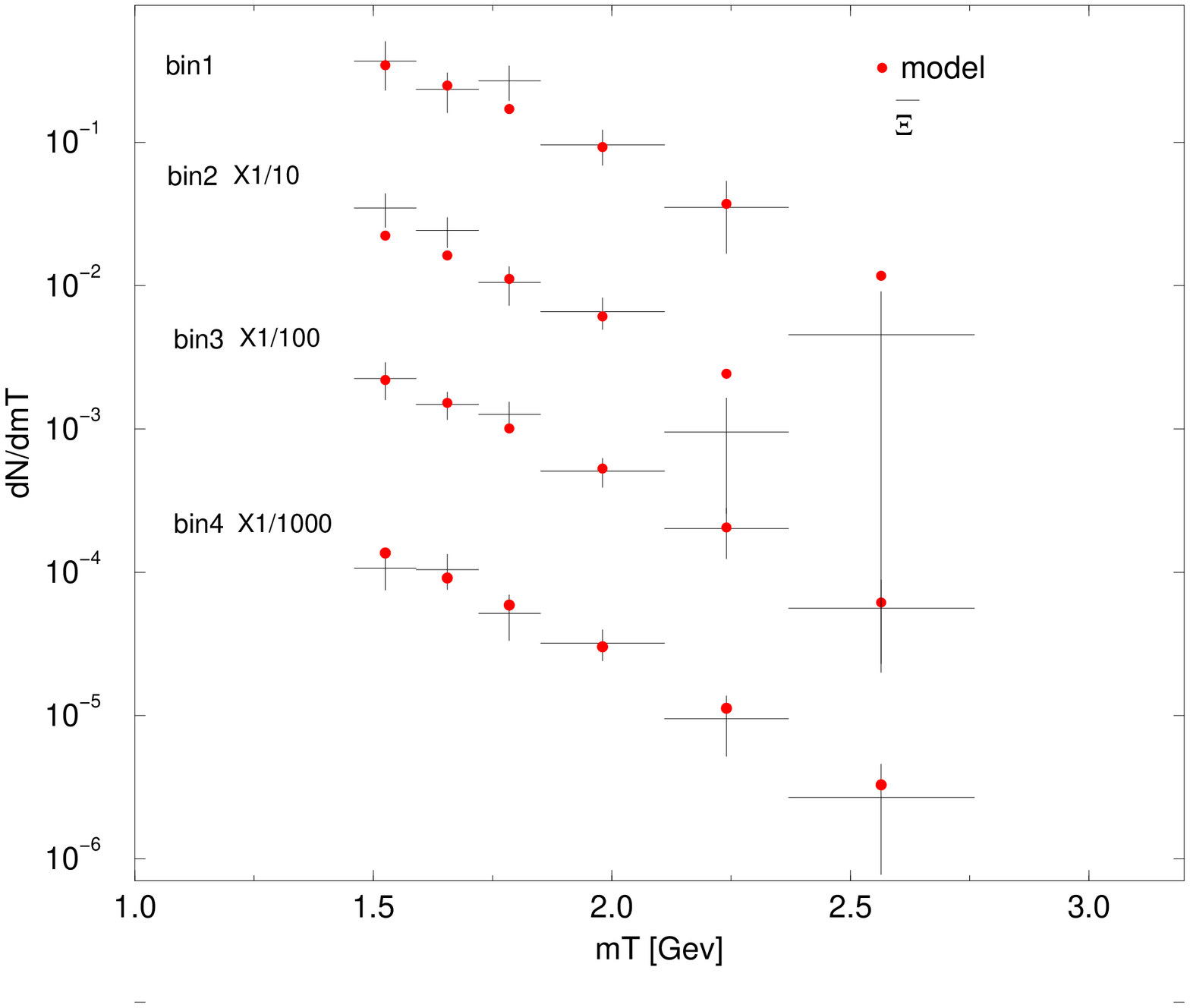}\\
\vspace*{0.3cm}
\end{tabular}
\caption{ 
Thermal analysis $m_T$ spectra: 
$\Lambda$ (top left), $\overline\Lambda$ (top right)
 $\Xi$ (bottom left),  $\overline\Xi$  (bottom right).
\label{TdLX}
}
\end{figure}
\begin{figure}[p]\hspace*{0.1cm}
\epsfig{height=7.cm,clip=,angle=-90,figure=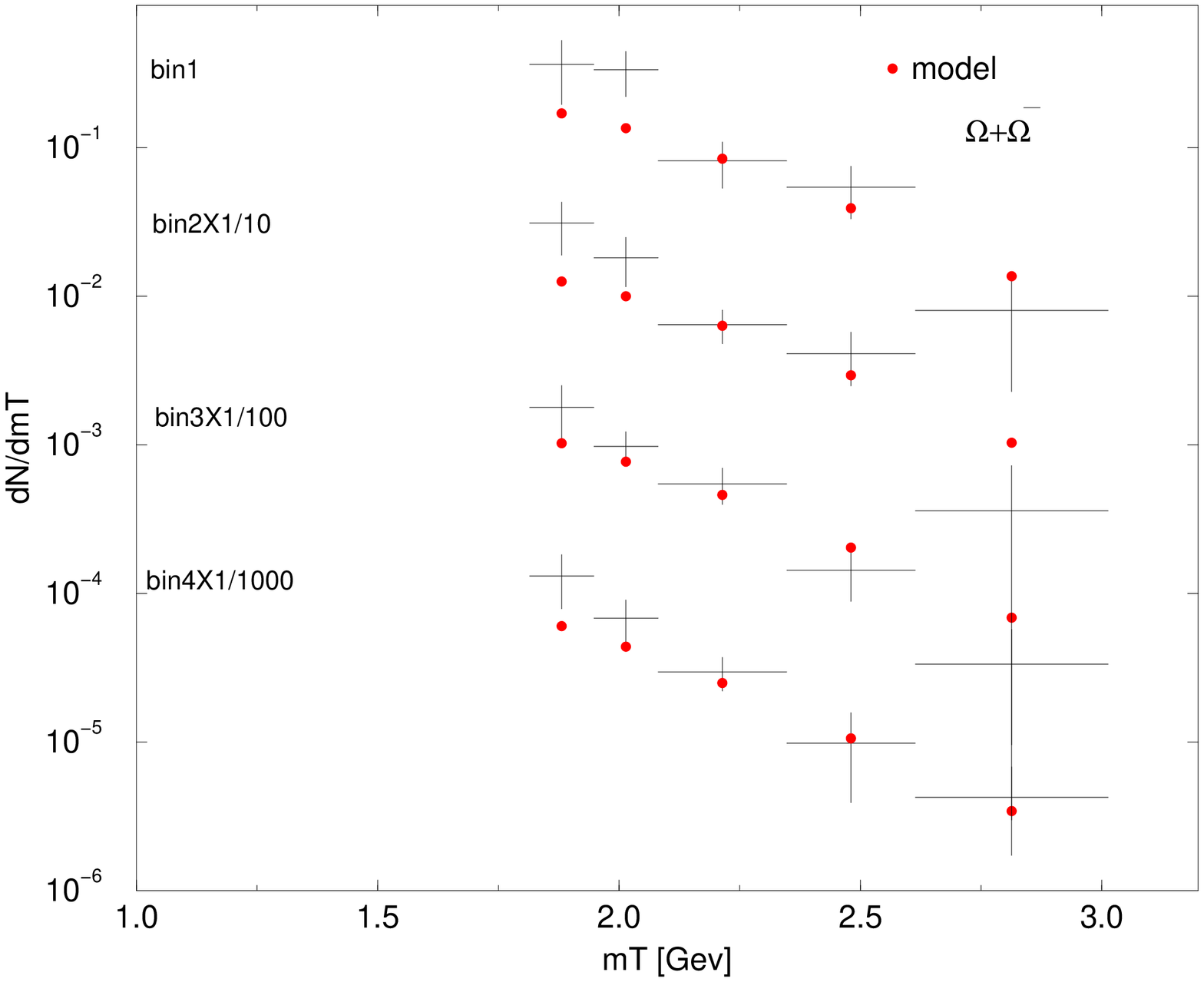}
\epsfig{height=7.6cm,clip=,angle=-90,figure=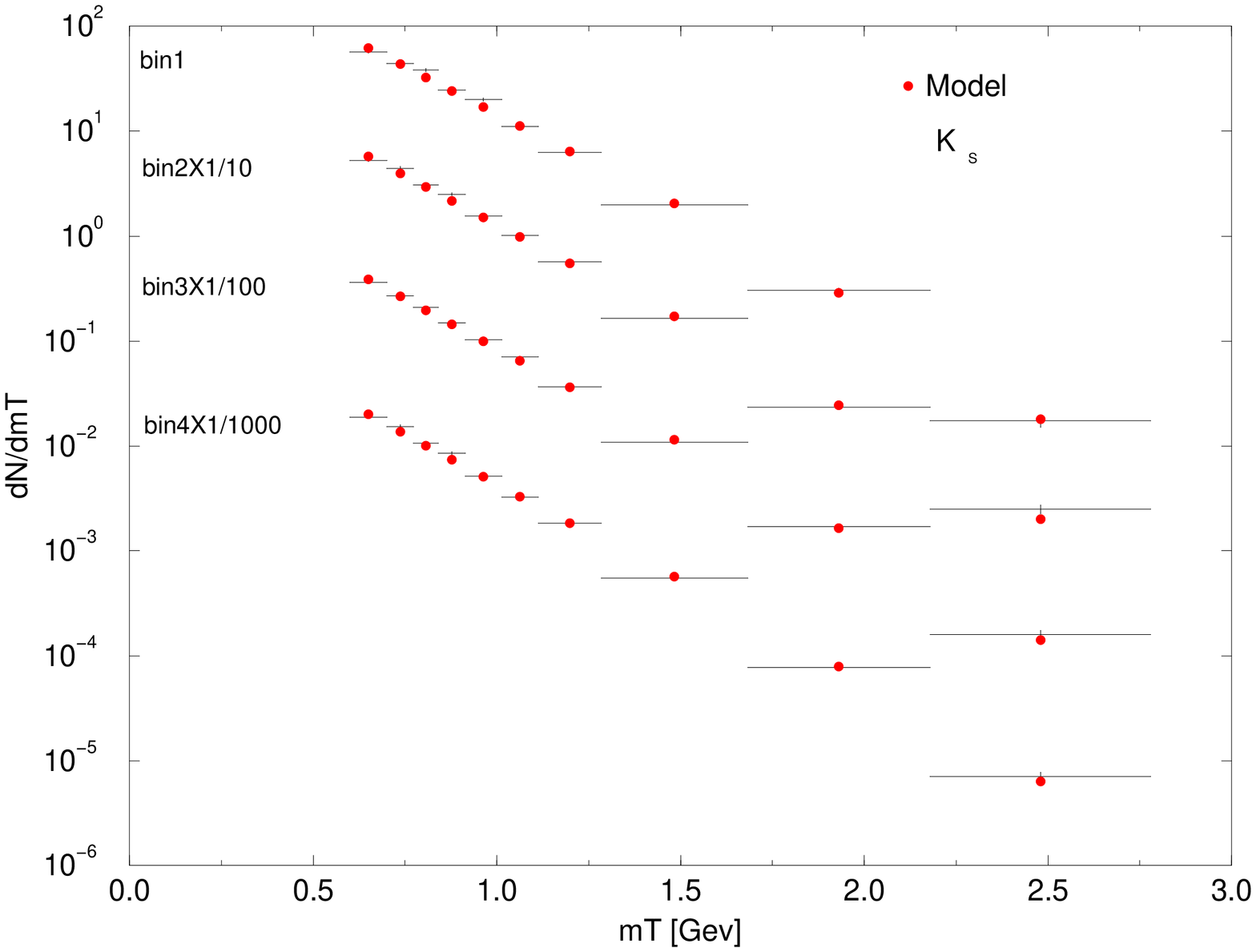}
\caption{ 
Thermal analysis $m_T$ spectra: $\Omega+\overline\Omega$ (left) and 
$K_{\rm s}$ (right).
\label{TdOK}
}
\end{figure}

It is important to explicitly check
how well the particle $m_\bot$-spectra are
reproduced. We group all centrality bin spectra and show, 
in figure \ref{TdLX},  $\Lambda,\,\overline\Lambda,\,\Xi$ and $\overline\Xi$. 
Overall, the description of the shape of the hyperon spectra 
is very satisfactory. In figure \ref{TdOK} on left we see for
the four centrality bins  the spectra for the sum $\Omega+\overline\Omega$.
The two lowest $m_\bot$ data points are systematically  under predicted. 
Some deviation at high $m_\bot$ may be attributable
to acceptance uncertainties, also seen in the the 
$\Xi$ results. This low $m_\bot$ enhancement of 
$\Omega+\overline\Omega$ is at the origin of the low value 
 of the inverse slope and the associated excess of $\Omega,\overline\Omega$
compared to the chemical freeze-out analysis. Unlike with chemical
analysis where the $\Omega,\overline\Omega$ have been omitted, 
the relatively large statistical errors allowed us
to include the  $\Omega$-spectra in the fit procedure, which
is dominated by the other hyperons and kaons. This allows to see
how the deviation from the systematics established by other
hyperons and kaons is arising.  In fact  
the low  $m_\bot$-bins of the $\Omega+\overline\Omega$ 
experimental spectrum with experimental yield excess at 1--2 s.d. 
translate into 3 s.d. 
deviations from the yields generated in the chemical analysis.

For kaons K$^{0}$ (figure \ref{TdOK} on right) the statistical errors are
very small, and we find in a more in depth statistical analysis that
they must be smaller than the systematic errors not considered. 
For this reason we have presented earlier in
figure \ref{TdTTdv1v2} results for temperatures obtained  with both
statistical, and  5 times greater than statistical error for kaons.
This increased value was used
in the fit with objective to asses the influence of the 
unknown systematic error. The stability of the result
implies that statistically precise kaons do not overwhelm the fit
procedure of hyperons, and/or that the hyperon and kaon
results are consistent with the model employed. 

Overall, we see that the description of kaon and hyperon spectra
with the parameters obtained in chemical freeze-out analysis if
 possible, indeed we conclude from
the above study that the thermal analysis is proving  the
conclusion that both thermal and chemical freeze-out occur
at the same condition of temperature and transverse velocity,
as one would expect in sudden break up of a deconfined supercooled 
quark--gluon plasma. Strange hadrons at SPS originate in single
freeze-out reaction.

\section{RHIC-130 hadron abundance analysis, round IIB}\label{final}
The strangeness yield observed in 158--200$A$ GeV reactions
corresponds to  $\cal O$(0.7)-$s\bar s$-pairs
of quarks per participant baryon. We estimate that this is 
corresponding to  50\%  QGP phase-space occupancy,
thus considerably more extreme results on strangeness 
can  arise at RHIC 
and with it greater anomalies in strange
baryon and antibaryon enhancement.
The experimental STAR and PHENIX collaboration
results considered here were  obtained  at $\sqrt{s_{\rm NN}}=130$ GeV
in the central-rapidity region, for the most central (5--7\%)
collision reactions.  The particle production 
results available as of
QM2002 meeting (Nantes, July 2002)  allow a
rather good understanding of the physical conditions 
established at the chemical hadron freeze-out at RHIC, and we use the 
opportunity to update our analysis.

Due to the approximate  
longitudinal scaling for the central rapidity results 
the effects of longitudinal flow at central rapidity cancels out and 
we can evaluate the full phase-space yields in order to obtain particle 
ratios. In our following analysis,  we do not include natural results
such as ${\pi}^+/{\pi}^-=1$, which can be expected 
since the large hadron yield combined with the
flow of baryon isospin asymmetry toward the fragmentation rapidity region 
assures this result  to within a great precision, as we
have discussed in section~\ref{Chem}. We also do not fit the results for
$\rm K^*$ and $\rm\overline{K^*}$ since the reconstructed yields depend on 
the degree of rescattering of resonance decay products~\cite{Mar02}. 
We consider here 19  particle ratios
seen in table \ref{RHIChad}.  We favor consideration of particle
ratios, since this  allows us to combine ratios from different
 experiments  with  slightly different trigger  conditions, and 
reduces  systematic errors. We have  
used published experimental  ratios and also formed ratios 
of rapidity densities reported by one and the same experiment.

We present in table \ref{RHIChad} in three last columns the results 
for both chemical equilibrium (last column) and non-equilibrium fits,
{\it i.e.} in the letter case we fix the chemical parameters to their equilibrium
value. We observe a considerable improvement in the statistical 
significance of the results of chemical non-equilibrium fits (see bottom line),
as we have seen in our related earlier RHIC work~\cite{Raf02a,Raf02b},
in consistency with the situation at SPS. Next to the fitted results,  
we show in parenthesis the contribution to the 
error ($\chi^2$) for each entry.
We consider  only  statistical errors for the experimental results, 
since much of the systematic error should cancel in the particle ratios. However, 
we do not allow, when pion multiplicity is considered, that errors  are  smaller
than $\simeq 8$\%, which is our estimated error in the theoretical evaluation
of the pion yield due to incomplete understanding of the high mass
hadron resonances. Some of the experimental results are 
thus shown with a theoretical error. 
When such an enlargement  of the experimental error
is introduced, a dagger as superscript appears in second column in 
table \ref{RHIChad} below. 

\begin{table}[!t]
\caption {\label{RHIChad}
Fits of central-rapidity hadron ratios at $\sqrt{s_{\rm NN}}=130$ GeV.
Top section: experimental results, 
followed in middle by chemical parameters (results of fits), and the
physical properties of the phase-space obtained from evaluation
of final state hadron phase-space, and the fitting error. 
Columns: ratio considered, data value with reference, the non-equilibrium
fit with 100\% $\Xi\to Y$ cascading $(f_\Xi=1)$ and 
40\% $Y\to N$ $(f_\Lambda=0.4)$, the non-equilibrium fit with 
40\% $\Xi\to Y$  and 40\% $Y\to N$, and in the last column, 
the chemical equilibrium fit with 40\% cascading . 
The superscript $^*$ 
indicates quantities fixed by constrains and related
considerations. The superscript $^\dagger$ indicates the error is dominated by
theoretical considerations. Subscripts $\Xi,\Lambda$ mean that these values
include weak cascading. In parenthesis we show  the contribution 
of the particular result to the total $\chi^2$.
}
\begin{tabular}{lccccc}
\hline\hline
  {\vphantom{$\frac AB$}}               &                 &      & 100\%\,$\Xi\to Y$&40\%\,$\Xi\to Y$ &40\%\,$\Xi\to Y$ \\
  {\vphantom{$\frac AB$}}               &Data             & Ref. & 40\% \ $Y\to N$  &40\%\,$Y\to N$   &40\%\,$Y\to N$  \\
\hline  
$ {\bar p}/p${\vphantom{$\frac AB$}}       &0.71 $\pm$ 0.06    &\cite{Phe02}&0.672(0.4) & 0.678(0.3)& 0.689(0.1)\\
${\overline\Lambda_\Xi}/{\Lambda_\Xi}$     &0.71 $\pm$ 0.04    &\cite{Sta02}&0.759(1.0) & 0.748(0.9)& 0.757(1.4)  \\
${\overline{\Xi}}/{\Xi}$                   &0.83 $\pm$ 0.08    &\cite{Cas02}&0.794(0.2) & 0.804(0.1)& 0.816(0.0) \\
$\rm{K^-}/{K^+}$                           &0.87 $\pm$ 0.07    &\cite{Phe01}& 0.925(0.6)& 0.924(0.6)& 0.934(0.8) \\
$\rm{K^-}/{{\pi}^\pm}$           & 0.15 $\pm$ 0.02$^\dagger$   &\cite{Phe01}& 0.159(0.2)& 0.161(0.3)& 0.150(0.0)  \\
$\rm{K^+}/{{\pi}^\pm}$           & 0.17 $\pm$ 0.02$^\dagger$   &\cite{Phe01}& 0.172(0.0)& 0.174(0.1)& 0.161(0.2) \\
${\Lambda_\Xi}/{h^-}$            & 0.059 $\pm$ 0.004$^\dagger$ &\cite{Sta02}& 0.057(0.3)& 0.050(5.1)& 0.045(11.9)  \\
${\overline\Lambda_\Xi}/{h^-}$   & 0.042 $\pm$ 0.004$^\dagger$ &\cite{Sta02}& 0.043(0.0)& 0.037(1.3)& 0.034(3.8) \\
${\Lambda_\Xi}/p$                          & 0.90 $\pm$ 0.12\  &\cite{Phe02}& 0.832(0.3)& 0.691(3.0)& 0.491(11.6) \\
${\overline\Lambda_\Xi}/\bar p$            & 0.93 $\pm$ 0.19\  &\cite{Phe02}& 0.929(0.0)& 0.763(0.8)& 0.539(4.2)  \\
$ \pi^\pm/p_\Lambda$                         &9.5 $\pm$ 2      &\cite{Phe01}& 9.4(0.0)  & 9.2(0.5)  & 7.6(22.8)  \\
$ \pi^\pm/{\bar p}_\Lambda$                  &13.4 $\pm$ 2.5   &\cite{Phe01}&13.7(0.1)  &13.4(0.0)  &10.9(7.9) \\
${\Xi^-}/{\pi}$                    &0.0088$\pm0.0008^\dagger$  &\cite{Cas02,QM02} &0.0096(1.0) & 0.0103(3.6) &  0.0067(7.1)  \\
${\Xi^-}/{h^-}$                    &0.0085$\pm$0.0015          &\cite{Cas02,QM02} &0.0079(0.1) & 0.0084(0.0) &  0.0054(4.3)  \\
${\overline{\Xi^-}}/{h^-}$         &0.0070$\pm$0.001           &\cite{Cas02,QM02} &0.0063(0.5) & 0.0068(0.1) &  0.0044(6.7)  \\
${\Xi^-}/{\Lambda}$                           &0.193$\pm$0.009 &\cite{QM02} &0.195(0.1) & 0.196(0.1) &  0.132(45.2)  \\
${\overline{\Xi^-}}/{\overline\Lambda}$       &0.221$\pm$0.011 &\cite{QM02} &0.213(0.6) & 0.214(0.4) &  0.144(48.7)  \\
${\Omega}/{\Xi^-}$                         & & &0.205 & 0.21 &  0.18 \\
${\overline{\Omega}}/{\overline{\Xi^-}}$   & & &0.22 & 0.23 &  0.20 \\
${\overline{\Omega}}/{\Omega}$             &0.95$\pm$0.1       &\cite{QM02} &0.87(0.7) & 0.88(0.5) &  0.89(0.4){\vphantom{$\frac AB$}} \\
${\bar p}/h^-$                             &                 &            &0.046     & 0.049      &  0.063   \\
${\phi}/K^-$                               &0.15$\pm$0.03    &\cite{QM02} &0.178(0.9)& 0.185(1.3) &  0.146(0.0)   \\
\hline  
$T${\vphantom{$\frac AB$}}     & &  &140.1 $\pm$ 1.1  & 142.3 $\pm$ 1.2    &164.3 $\pm$ 2.2   \\
$\gamma_{ q}^{\rm HG}$         & &  &1.64$^*$         & 1.63$^*$           &  1$^*$            \\
$\lambda_{ q}$                 & &  &1.070 $\pm$ 0.008& 1.0685 $\pm$ 0.008 &1.065 $\pm$ 0.008\\
$\mu_b$ [MeV]                  & &  &28.4             &28.3                &31.0                 \\
$\gamma_{ s}^{\rm HG}/\gamma_{ q}^{\rm HG}$ & &  &1.54 $\pm$ 0.04   &  1.54 $\pm$ 0.0.04    &  1$^*$               \\
$\lambda_{ s}$                 & &  &1.0136$^*$       &  1.0216$^*$        & 1.0196$^*$            \\
$\mu_{\rm S}$ [MeV]            & &  &6.1              &6.4                 &7.1                    \\
\hline{\vphantom{$\frac AB$}}
$E/b$ [{\small GeV}]    $\!\!$ & &  &35.0             &34.6                &  34.8    \\
$s/b$                          & &  &9.75             &  9.7               &  7.2      \\
$S/b$                          & &  &234.8            & 230.5              & 245.7     \\
$E/S$ [{\small MeV}]   $\!\!$  & &  &148.9            & 150.9              & 141.5       \\
\hline    
$\chi^2/$dof                   & &  &7.1/($19-3$)     &19/($19-3$)        &177.2/($19-2$) \\
\hline\hline
\end{tabular}
\end{table}
As is today well understood, the high yields of hyperons
require significant corrections for unresolved weak decays. 
Some experimental results are already corrected in this fashion:  the weak
cascading corrections were applied to the most recent $p$ and $\bar p$ 
results by the PHENIX collaboration \cite{Phe02}, and in the $\Xi/\Lambda$
and $\overline\Xi/\overline\Lambda$ ratio of the STAR collaboration we
use here \cite{Cas02,QM02}. However, some of the results
we consider are not yet corrected \cite{Phe01,Sta02}, and are indicated 
in the first column in table \ref{RHIChad} by a subscript $\Lambda$ or $\Xi$. 

The  subscript $\Lambda$ means 
that the cascading of hyperons into nucleons has to be included while
fitting the particle ratio considered, and we have found that a 35-40\% cascading 
($f_\Lambda =0.35$-0.4) is
as expected favored  by the fits of pion to nucleon ratio.
Similarly,  the subscript $\Xi$ indicates 
that the $\Xi$ cascading  into single strange hyperons $Y=\Sigma,\Lambda$
was also not corrected for in the considered result. 
Here we find empirically that full acceptance of the Cascades into 
singly strange hyperons is favored, again   by both equilibrium and 
non-equilibrium fits ($f_\Xi\to 1$). We also
include in the $\Xi$ and $\Lambda$ yield with the appropriate branching the 
weak decay of $\Omega$, with the experimental acceptance fraction 
 being the same as in the decay of $\Xi$  ($f_\Omega=f_\Xi$).

In other words, in the results shown in table \ref{RHIChad},  
the hyperon yields are:
\begin{eqnarray*}
\Lambda_\Xi&=&\Lambda_{\rm th}+\Sigma^0_{\rm th}
   +f_\Xi(2\Xi^-_{\rm th}+\Omega_{\rm th}),\quad 
\overline\Lambda_\Xi=
   \overline\Lambda_{\rm th}+\overline\Sigma^0_{\rm th}+
 f_\Xi(2\overline{\Xi^-_{\rm th}}+\overline\Omega^-_{\rm th});\\[0.2cm]
p_\Lambda&=& p_{\rm th}
   +f_\Lambda 0.77 \Lambda_\Xi, \qquad\qquad\quad\,\   
\bar p_\Lambda= \bar p_{\rm th}
   +f_\Lambda 0.77 \overline{\Lambda_\Xi};\\[0.2cm]
\Xi^-&=&\Xi^-_{\rm th}+0.08\Omega^-_{\rm th},\qquad\qquad\quad\quad\,
\overline{\Xi^-}=\overline{\Xi^-_{\rm th}}+0.086\overline{\Omega^-_{\rm th}}.
\end{eqnarray*}
Here subscript `th' indicates statistical model yields. The factor `2' preceding
$\Xi^-_{\rm th},\overline{\Xi^-_{\rm th}}$ allows for the decay of  $\Xi^0,\overline{\Xi^0}$
which are assumed to be equally abundant. However, there is considerable isospin asymmetry in
the decay pattern, and the branching ratio 
weight 0.77 arises as follows: 64\% of $\Lambda$ and 51.6\% 
of $\Sigma^+$  decay to protons. Statistical model evaluation shows that  $\Sigma^+$
(the only isospin channel decaying into $p,\bar p$)
are produced at the level of 25\% of $\Lambda+\Sigma^0$ (which one usually calls 
$\Lambda$),  considering within the statistical model that a significant   
fraction of all $\Lambda$  originates in the $\Sigma^*(1385)$ resonance.

Below the fit  results we show the statistical parameters  which are related to each fit. 
The results shown  in the table \ref{RHIChad} are 
obtained minimizing in the space of 3 parameters: the chemical freeze-out temperature
$T$, and 2 chemical parameters $\lambda_q,\gamma_s$, the value of 
$\gamma_q$ is set at its maximal value $\gamma_q^2={\gamma_q^c}^2=e^{m_\pi/T}$
and the value of $\lambda_s$  is derived from the 
strangeness conservation constraint. We note that even without
this requirement the  fit converges to local strangeness neutrality within a 
few percent.

In table \ref{RHIChad} the last column presents the results of chemical
equilibrium fit with 40\% cascading. We note that several particle yileds
are not properly described and hence a large $\chi^2$ results. However, 
on a logarithmic scale only results involving $\Lambda,\overline\Lambda$ would
be clearly visible as a discrepancy. The second and third last column 
show result of chemical non-equilibrium fit, the second last with 40\% cascading 
which yields a good fit, and the third last with an increased 100\% acceptance 
for the $\Xi$-cascading gives the smallest $\chi^2$, and has a very high
 confidence level.

Our current  results, when compared to
our earlier effort~\cite{Raf02a,Raf02b} show a 7\%
reduction in the freeze-out temperature
both for equilibrium and non-equilibrium case. Such lower $T$
  reduces selectively the 
relative yield of  baryons. The required level of 
proton and hyperon  yields  are now reached at lower temperature
due to the introduction
of strange baryon weak cascading, or respectively, of experimental results  which 
are corrected for cascading.  The range of  
temperature now seen agrees better with the expectations we had 
considering the effect of the fast expansion  of QGP~\cite{Raf00}. 
The chemical equilibrium freeze-out temperature, $T=164$\,MeV,
is in  agreement with results of chemical equilibrium single freeze-out 
model of  Broniowski and Florkowski \cite{Bro01,Bro02}.
We believe that the case for single freeze-out particle production at RHIC
 will be statistically stronger  with introduction
of chemical non-equilibrium. We have seen this happen 
in section \ref{thermal} for the  SPS results.

Returning to discuss results seen in the bottom of table~\ref{RHIChad} we note
that the specific strangeness content $s/b\simeq 10$ 
in chemical non-equilibrium fits is 35\% greater than the
chemical equilibrium result. This originates in 
$\gamma_s/\gamma_q\simeq 1.5$ ({\it i.e.} $\gamma_s\simeq 2.5$). 
Figure \ref{PLSBLAMQ} ahows that  the chemically equilibrated
QGP phase-space  has   10-14 pairs of strange quarks per baryon
at $1.062<\lambda_{ q}<1.078$. This implies  that  at RHIC-130 we have
$\gamma_{ s}^{\rm QGP}\simeq 0.85\pm0.15$.  This in turn {\it theoretically}
implies that  $\gamma_{ s}^{\rm HG}\simeq ${\bf 3}$ (0.85\pm0.15)=2.5\pm5$ given the
enhancement of the HG phase space occupancy by a factor three 
seen in  figure \ref{PLGAHGQGP}. Thus the specific yield
 of strangeness and strangeness occupancy we measure in the HG 
after hadronization is consistent with  QGP properties with nearly 
saturated strangeness phase space. Even so, a further enhancement of $\gamma_s$ 
can be expected in the 200 GeV RHIC run. 

Once the comparison $N$-$N$ experimental
results will become available, the enhancement of the production of
multistrange baryons by this extraordinary strangeness abundance must
exceed the high values seen at SPS by a factor $(1.5$-$2.5)^{n_s}$, where
$n_s$ is strangeness content of the hadron. An enhancement by a factor 
50--300 of the $\Xi,\,\overline\Xi,\,\Omega,\,\overline\Omega$ yields 
is  difficult to explain without new physics,
and is natural for the fast hadronizing baryonpoor QGP phase. 


In closing we would like to emphasize that  the chemical 
non-equilibrium description of hadronization process  
is statistically highly favored at SPS and RHIC.
Both the particle spectra and the HBT measurement of space-time size
are  favoring the sudden hadronization of a relatively small, short lived
and rapidly  expanding matter fireball, which reaction picture
necessitates  presence of chemical non-equilibrium. 
We hope that the reader who studies this report gains further the
impression that a non-equilibrium chemical analysis of heavy-ion particle
yields offers profound insight into the physical properties of the
dense hadronic matter formed in the relativistic heavy-ion collisions. 

\begin{theacknowledgments}
Work supported in part by a grant from the U.S. Department of
Energy,  DE-FG03-95ER40937, and by NSF grant INT-0003184. 
Laboratoire de Physique Th\'eorique 
et Hautes Energies, LPTHE, at  University Paris 6 and 7 is supported 
by CNRS as Unit\'e Mixte de Recherche, UMR7589.
\end{theacknowledgments}


\bibliographystyle{unsrt3}   

\bibliography{Rafbookt}

\IfFileExists{\jobname.bbl}{}
 {\typeout{}
  \typeout{******************************************}
  \typeout{** Please run "bibtex \jobname" to obtain}
  \typeout{** the bibliography and then re-run LaTeX}
  \typeout{** twice to fix the references!}
  \typeout{******************************************}
  \typeout{A faire deux fois}
 }

\end{document}